\documentclass[reprint,
 amsmath,amssymb,
 s,ap
 pre,
]{revtex4-1}
\usepackage{graphicx}
\usepackage{dcolumn}
\usepackage{bm}
\usepackage{float}
\usepackage{subfigure}
\usepackage[abs]{overpic}
\usepackage{hyperref}%
\usepackage{lineno}
\usepackage[dvipsnames]{xcolor}
\usepackage{titlesec}
\usepackage{soul}
\usepackage{multirow}
\usepackage{array}
\usepackage{tabularx}
\usepackage{makecell} 
\usepackage{xr}

\newcommand{\bb}{\textcolor{blue}}

\begin{document}
\makeatletter
\newcommand*{\rom}[1]{\expandafter\@slowromancap\romannumeral #1@}
\makeatother
\preprint{APS/123-QED}

\title{Topological trapping of active tracers in branched ring polymer matrices}

\author{Rahul Bag}
\affiliation{Department of Physics, National Institute of Technology, Warangal, 506004, India}

\author{Baburao Simma}
\affiliation{Department of Physics, National Institute of Technology, Warangal, 506004, India}

\author{Vrinda Garg}
\affiliation{Department of Physics, National Institute of Technology, Warangal, 506004, India}

\author{Surya K. Ghosh}
\email{skghosh@nitw.ac.in}
\affiliation{Department of Physics, National Institute of Technology, Warangal, 506004, India}

\date{\today}

\begin{abstract}
Branched macromolecular architectures are ubiquitous in living systems and synthetic polymeric systems, create topologically complex media that strongly influence tracer dynamics beyond simple packing fraction-driven crowding effects. Motivated by biological systems such as amoebae, astrocytes, cytoskeletal assemblies, and cyclic bottlebrushes, we investigate active and passive tracer transport in polymeric environments of topologically distinct architectures. Here, using Langevin dynamics simulations, we focus on how polymer topology alters diffusion behavior in a branched-ring polymeric (BRP) environment. To isolate the role of intramolecular topology from monomer density, we specifically benchmark the BRP dynamics against an unlinked linear-ring polymer mixture (LRM). In general, we find that while passive tracers are largely unaffected by the polymer connectivity and bending rigidity, the self-propelled active tracers strongly distinguish between the two architectures. Increasing chain rigidity facilitates active diffusion in the unlinked LRM but strongly restricts mobility in the BRP matrix, driving a transition to pronounced long-time subdiffusion that becomes more prominent with propulsion strength. The ergodicity-breaking parameter, van Hove correlation function, and non-Gaussian parameter reveal that interconnected topological cages are formed in the stiff limit, leading to transient weak ergodicity breaking and leptokurtic displacement statistics. Further, the self-intermediate scattering function confirms that the short-range dynamics remain unaltered by the matrix connectivity, while relaxation at large spatial length scales is governed by the topological caging. Finally, varying the branch length and number, we identified an optimal branch architecture that maximizes active trapping. Our results establish that macromolecular connectivity and stiffness dominate the active transport even in the system of identical packing fraction.

\end{abstract}

\maketitle

\section{Introduction}

The transport of tracers through crowded polymeric media is central to intracellular trafficking, signaling \cite{kozaeva2025getting}, drug delivery \cite{Ralf_PRE_2024, Liu_ChiChemLet_2023, Jeon_MacMol_2022, Jeon_MacMol_2024, roy2021kinetics}, and materials science \cite{Chen_JoR_2021, Zhao_PCCP_2022, Yuan_JCP_2016, Fang_MatToBio_2023, Composto_MacMol_2011, Mooney_NatRevMater_2016, Ralf_SoftMatter_2019}. Tracer motion in polymeric media is governed not only by thermal fluctuations but also by viscoelasticity, crowding, and crucially, the topology of the surrounding polymeric medium \cite{Fytas_PRL_2013, Li_PRE_2025, Zhao_PCCP_2019}. Consequently, tracer dynamics often deviates strongly from classical Brownian diffusion, exhibiting anomalous transport \cite{garg2024crowding, garg2026semiflexibility} and caging behavior. Many biologically relevant systems comprise active particles that consume energy to sustain persistent motion \cite{Aranson_ACR_2018, Doostmohammadi_NatCom_2018}. Molecular motors \cite{Liu_ScienceAdv_2023}, swimming bacteria \cite{Schwarz_PNAS_2012}, and self-propelled colloids \cite{Wang_ACR_2015, Baskaran_CurOpCellBio_2016} are prominent examples whose directed motion differs fundamentally from the thermal diffusion of passive particles \cite{Ramaswamy_RevModPhys_2013, Bechinger_Review_2016, Menon_RheoComFlu2010}. When an active tracer moves through a polymeric environment, its persistent propulsion can couple strongly to the local structure and dynamics of the surrounding polymers, leading to transport behavior that differs fundamentally from that of a passive Brownian particle \cite{Hou_SoftMatter_2019, Zhao_PCCP_2022, Zhao_PCCP_2019, Li_PRE_2025}.

Among the various factors that play a decisive role in tracer dynamics, the topology arising from the macromolecular connectivity of the surrounding polymeric medium is crucial \cite{Chen_JoR_2021, Hou_SoftMatter_2019, Mukhopadhyay_MacMol_2012, Li_PRE_2025, Jeon_MacMol_2022, Jeon_MacMol_2024}. For example, ring polymers, which lack free ends, exhibit compact conformations and unique topological constraints compared with linear polymers \cite{Gupta_MacMol_2017, Morgese_2016, Chen_MacMol_2023}, which directly alter the tracer dynamics \cite{Chen_JoR_2021, Zhao_PCCP_2022, Jeon_MacMol_2022, Jeon_MacMol_2024}. However, beyond simple linear and ring polymer architectures, biological and synthetic polymeric systems consist of branched macromolecules of more complex connectivity \cite{Jeon_SoftMatter_2020}. These introduce structural heterogeneity and topological constraints that are absent in both linear and ring polymers, and that cannot be captured by monomer density alone. Understanding the influence of such environments in tracer transport remains largely unexplored.

Within this broad context, branched macromolecules constitute a fundamental and ubiquitous class of soft matter systems, particularly in biological environments \cite{Kurt_Binder_JCP_1991, Scott_Biochemical_1991, Liman_2020, Inoue_ProPolSci_2000, Haydukivska_SciReports_2020, Blavatska_2015, Tian_PhyLetA_2019, Roichman_JCP_2021, Pugliese_ABiomaterialia_2018}. Examples range from chromatin with looped and branched domains \cite{Bohn_JCP_2010, Matthew_JBioMatRes_2007},  glycogen and amylopectin networks to cytoskeletal assemblies and intracellular polymeric scaffolds \cite{Bohn_macromoelcules_2010}. Such architectures are not incidental; rather, branching plays a crucial role in regulating mechanical stability, transport, and dynamic reorganization in living systems \cite{Bebber_2007, Zaferani_2026, Blanchoin_PhyReview_2014, Liman_2020}. Beyond biology, branched polymers are of great importance in medicine and materials science, and have long been a subject of interest in polymer physics, due to their tunable rheology, enhanced solubility, unique transport, and relaxation properties \cite{Liman_2020, Zaferani_2026, Pugliese_ABiomaterialia_2018}.

Grafting linear branches onto a closed cyclic backbone produces branched ring polymers (BRP), a distinct architectural class that combines closed loop with open-ended side-chain connectivity, a unique topological constraint absent in either linear, ring, or purely branched polymers. Such branch-ring polymers (BRPs) with hybrid topologies are not only of fundamental interest but are also realized in synthetic systems, including cyclic brush \cite{Zhang_JACS_2013, Morgese_2016, Shi_MacmolRapCom_2023} and cyclic bottlebrush polymers \cite{Pal_MacMol_2022, Shi_MacmolRapCom_2023, Chen_MacMol_2023}, whose unique mechanical and transport properties have attracted increasing attention. To the best of our knowledge, transport through a branched-ring polymeric environment is limited experimentally and computationally.

Beyond synthetic materials, similar looped and branched architectures are ubiquitous in biological systems, such as the dynamic actin cytoskeleton of amoebae \cite{Blanchoin_PhyReview_2014, Roichman_JCP_2021}, which continuously reorganizes into branched and looped networks during migration and phagocytosis, and the arborized morphology of central nervous system astrocytes \cite{Blanchoin_PhyReview_2014, Liman_2020, Baldwin_TreCellBio_2023, Wynne_bioRxiv_2025, Viana_2023}. In these media, diffusing tracer particles encounter a hierarchical assembly of loops and branches rather than a homogeneous linear polymer solution or a randomly crosslinked gel. 

While mixtures of unlinked linear and ring polymers (LRM) have been studied extensively \cite{RosaAngelo_softmatter_2021}  and provided valuable insights into topologically mediated tracer diffusion, a BRP environment is intrinsically different, as its topological constraints are built intramolecularly into the molecular architecture itself. This distinction raises a fundamental question: how does tracer diffusion in BRP systems differ from that in unlinked LRM, and to what extent does polymer topology dominate over conventional crowding parameters such as packing fraction or monomer density, which alone cannot capture the constraints generated by branching and loop closure? While packing fraction has traditionally been used as a primary descriptor of crowding and anomalous diffusion, growing evidence suggests that topology can play an equally critical role. A topology-focused approach is therefore essential for understanding transport in such complex environments. 

In this work, we use two-dimensional Langevin dynamics simulations to investigate the transport of both passive (Brownian) and active (self-propelled) tracers in polymeric environments composed of BRP and compare the results with those obtained in LRM. Our study aims to isolate the role of topology from that of packing fraction. To be precise here, we explicitly investigate tracer transport in branched ring polymeric environments, and we use LRM as a reference system to isolate the effects of complex connectivity in polymer architecture and topology. By systematically contrasting passive and active transport, we elucidate how non-equilibrium driving interacts with topological constraints arising from branching and loop closure. Our results demonstrate that tracer transport is governed by an interplay between activity, polymer stiffness, and polymer topology. Passive tracers are slightly sensitive to the stiffness or architecture, whereas active tracers strongly distinguish between the BRP and LRM environments. In particular, stiff BRP systems exhibit pronounced long-time subdiffusion, reduced diffusivity, stronger transient weak-ergodicity breaking, and enhanced leptokurtic non-Gaussian displacement statistics, reflecting heterogeneous transport characterized by persistent motion interspersed with intermittent topological trapping. The self-intermediate scattering function further confirms that these architectural effects govern structural relaxation at large length scales, while short-range diffusion remains dictated by local monomer crowding.

We also find a non-monotonic dependence of tracer diffusivity on the branch geometry of BRPs at a fixed packing fraction, with an intermediate branch length and number identifying an optimal architecture for active trapping. Taken together, these results establish that polymer packing fraction alone is insufficient to characterize tracer transport: the connectivity, stiffness, and spatial organization of the polymer matrix qualitatively alter active-particle dynamics, with activity itself driving the tracer into topological traps. These results provide insights relevant to biological transport, targeted drug delivery systems, and the rational design of advanced synthetic smart or functional soft materials where branched macromolecular architectures are the rule rather than the exception\cite{Guironet_MacroMol_2026}.

The remainder of the paper is organized as follows. In Section II, we describe the model and simulation methods. Physical observables are represented in Section III. Section IV presents and discusses our main results. Finally, Section V summarizes the key findings and presents the main conclusions of the paper.

\begin{figure}
\centering
\includegraphics[width=8cm]{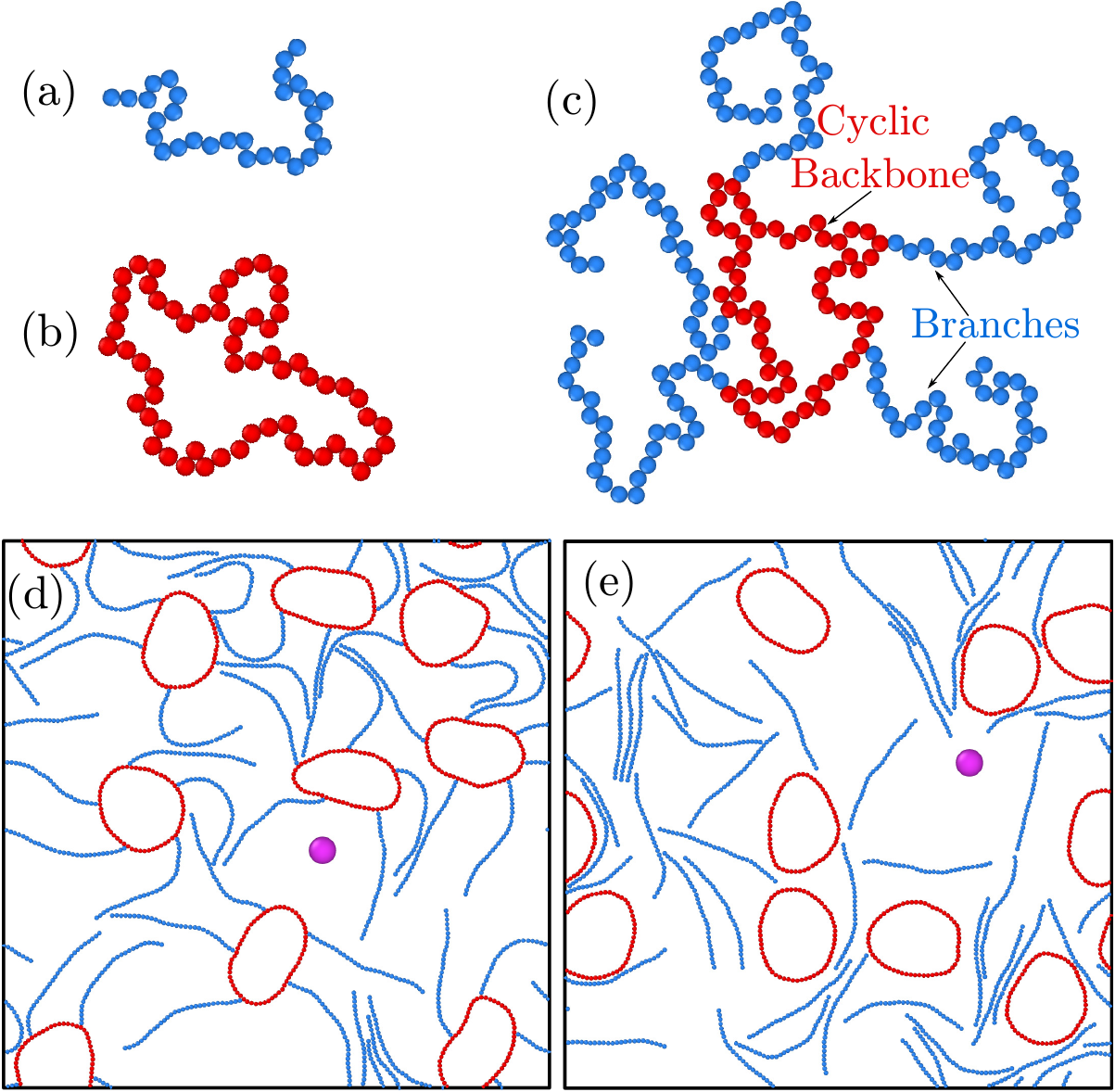}
\caption{Schematic representation of a flexible ($\kappa = 0$) (a) Linear Polymer (LP), (b) Ring Polymer (RP), and (c) Branched Ring Polymer (BRP), which consists of five branch polymers (BP) connected to a Cyclic Backbone (CB). Representative simulation snapshots of the stiff ($\kappa = 100$) (d) Branched Ring Polymer (BRP) matrix and (e) unlinked Linear-Ring Mixture (LRM) environment, maintained at an identical packing fraction ($\phi = 0.1$). Note that the active tracer (magenta) is enlarged for better visualization; for all calculations and analysis, the diameter of the tracer is taken to be the same as the monomers.}
\label{Schematic}
\end{figure}

\section{Model and Methods}
\label{sec-model}
We investigate the dynamics of a self-propelled tracer particle in topologically distinct polymeric environments using two-dimensional Langevin dynamics simulations. In our work, the tracer particle is modeled as an active Brownian particle (ABP) whose intermediate-time dynamics is non-Gaussian in free space. The first environment comprises Branched Ring Polymers (BRP), whereas the second is a Linear and Ring Polymer Mixture (LRM)(see Fig. \ref{Schematic}; schematics in (a, b \& c) and system snapshots in (d, e)). Each BRP consists of a cyclic backbone (CB) grafted with equally spaced linear polymer branches (Fig. \ref{Schematic}(c)). We denote the number of branches in a single BRP by $m_b$, the number of BRPs in the system by $m_c$, and the number of monomers in the cyclic backbone and each branch by $n_c$ and $n_b$, respectively. Here, $n_b$ corresponds to branch length. Thus, the total number of branches in the BRP system is $M_b=(m_b \times m_c)$, while the number of monomers in a single BRP is $N_b=n_c+(m_b \times n_b)$. In our system, $N_b = 210$. The LRM comprises $m_r$ ring polymers (RP) and $M_l$ linear polymers (LP), where each RP contains $n_r$ monomers, and each LP contains $n_l$ monomers, which corresponds to the length of a linear polymer. To facilitate a systematic comparison between the two topologically distinct environments, we match the corresponding structural parameters according to $m_c=m_r=9$, $M_b=M_l$, $n_c=n_r=60$, and $n_b=n_l=30$; see TABLE~\ref {table_para_rel}.

The total number of monomers in the BRP and LRM systems is given by

\begin{equation}
\begin{aligned}
N & = m_c (n_c + (m_b\times n_b)), && \textit{in}~~~~\text{BRP}\\
  & = (m_r\times n_r) + (M_l\times n_l), && \textit{in}~~~~\text{LRM}
\end{aligned}
\label{total_monomers}
\end{equation}

Consequently, both systems contain the same total number of monomers, $N=1890$, which is kept fixed in both systems throughout the study.

The active tracer and all polymer beads have the same diameter, $\sigma=1$. Excluded volume interactions between monomer-monomer and monomer-tracer are purely repulsive, modeled by the Weeks-Chandler-Andersen (WCA) potential,

\begin{equation}\label{LJ_pot}
U_ {\text{LJ}} (r) = 4\epsilon\left[\left(\frac{\sigma}{r}\right)^{12} - \left(\frac{\sigma}{r}\right)^{6} + \frac{1}{4}\right]\times\Theta(r_{\text{cut}}-r).
\end{equation}

Here, $r$ is the interparticle separation, $\epsilon = 1$ defines the uniform interaction strength, and $\Theta$ is the Heaviside step function that truncates the attractive part of the LJ potential at $r_{\text{cut}}=2^{1/6}\sigma$. The purely repulsive potential ensures the excluded volume effect and self-avoidance between the monomers and the tracer particle. Adjacent monomers within the polymer chains are bonded via the finitely extensible nonlinear elastic (FENE) potential,

\begin{equation}
\label{fene_eq1}
U_{\text{FENE}}(r)
=-\frac{k_fR_0^2}{2}
\ln\left(1-\frac{r_{i,i-1}^2}{R_0^2}\right),
\quad r_{i,i-1}<R_0.
\end{equation}

where $r_{i,i-1}=|r_i-r_{i-1}|$ represents the separation between consecutive monomers $i$ and $i-1$. The spring constant is set to $k_f = 30\epsilon/\sigma^2$\cite{Luo_PCCP_2018}, and $R_0 = 1.5\sigma$ denotes the maximum allowed bond extension. The bending potential between consecutive bonds is represented as \cite{Luo_PCCP_2018},

\begin{equation}\label{bend_eq2}
U_{\text{BEND}}(\varphi)=\kappa(1 + cos\varphi)
\end{equation}

where $\kappa$ represents the bending rigidity and $\varphi$ is the angle between two consecutive bond vectors.

 We define the stiffness of different components of the system separately: the bending rigidity of CB and BP in the BRP system are denoted by $\kappa_{c}$ and $\kappa_{b}$, respectively. While in the LRM system, the corresponding parameters are $\kappa_{r}$ (RP) and $\kappa_{l}$ (LP). By varying the bending rigidity from fully flexible ($\kappa = 0$) to highly semi-flexible or stiff ($\kappa = 100$), we systematically investigated the whole parameter space. To ensure the topological differences, we mapped the respective bending rigidities $\kappa_{c} = \kappa_{r}$ and $\kappa_{b} = \kappa_{l}$.

The time evolution of the position vector $\bm{r_{i}}(t)$ of the $i^{th}$ monomer of the polymer is governed by the overdamped Langevin equation,

\begin{equation} \label{Eq:Langevin_Polymer}
\gamma \frac{d\boldsymbol{r}_i}{dt} = -\boldsymbol{\nabla}_i U_{\text{tot}} +\boldsymbol{f}_i(t),
\end{equation}

where $U_{tot}= U_{LJ} + U_{FENE} + U_{BEND}$, is the total interaction potential, $\gamma$ is the friction coefficient, and $f_i(t)$ is the Gaussian white noise with zero mean and the autocorrelation following the fluctuation-dissipation theorem given by $\langle \boldsymbol{f}_i(t) \cdot \boldsymbol{f}_j(t') \rangle = 4 \gamma k_B T \delta_{ij} \delta(t-t')$. Here $k_B$ is the Boltzmann constant and $T$ is the temperature of unit magnitude. 

The dynamics of the active particle at a position $\bm{r_a(t)}$ is described by the overdamped Langevin equation as  \cite{Romanczuk_EJP_2012},

\begin{equation} \label{Eq:Langevin_Particle}
\gamma \frac{d\boldsymbol{r}_a}{dt} = -\boldsymbol{\nabla}_a U_{\text{LJ}} + \boldsymbol{f}_a(t) + F_a \boldsymbol{n},
\end{equation}

where $f_a(t)$ , is the translational Gaussian white noise with $\langle f_a(t) \rangle = 0$, follows the fluctuation-dissipation theorem with the translational diffusion coefficient, $D_T = k_BT/\gamma$. The constant self-propulsion force $F_a$, acts along the unit vector $\boldsymbol{n}$, which takes random orientation with time as \cite{Hou_SoftMatter_2019}

\begin{equation} \label{Eq:dn_dt}
\frac{d\boldsymbol{n}}{dt}= \boldsymbol{\eta}(t) \times \boldsymbol{n}
\end{equation}

where vector $\boldsymbol{\eta(t)}$ is a Gaussian white noise in cartesian co-ordinate. It is characterized by zero mean and a correlation of $\langle\boldsymbol{\eta} (t) \cdot \boldsymbol{\eta}(t')\rangle = 2D_R\delta(t-t')$. The rotational diffusion constant $D_R$ is governed by the persistence time of directional reorientation, and is coupled to the translational diffusion via $D_R = 3 D_T /\sigma^2$. In two dimensions, the random vector has only z-component rotation i.e. $\eta(t) = (0, 0, \eta_z(t))$. Eq. \ref{Eq:dn_dt} in plane polar form $(cos\theta, sin\theta, 0)$\cite{Lowen_JCP_2020} thus reduced to

\begin{equation} \label{unit_vector_nx_and_ny}
\frac{d\boldsymbol{n_x}}{dt} = -\eta_z sin\theta, \frac{d\boldsymbol{n_y}}{dt} = \eta_z cos\theta.
\end{equation}

Consequently, the rotational change in the unit vector is given by $\frac{d\theta}{dt} = \eta_z(t)$\cite{Lowen_JCP_2020}. In our work, we quantified the active strength by a non-dimensional parameter, i.e. P\'{e}clet number, $Pe = \frac{F_a\sigma}{k_BT}$. It is the ratio between active and diffusive transport \cite{Bechinger_Review_2016, Jeon_MacMol_2022}. Here, $Pe=0$ corresponds to a passive particle in a passive polymeric system, whereas $Pe > 0$ denotes an active self-propelled particle. 

In our simulation, as we specifically focus on the topological effect, both the BRP and the reference LRM systems are maintained at an identical packing fraction. The equations of motion are integrated using a velocity-Verlet algorithm within LAMMPS \cite{PLIMPTON_LAMMPS}. We impose periodic boundary conditions across a $120\sigma \times 120\sigma$ simulation box. To approximate the overdamped limit characteristic of polymeric suspensions, a Langevin thermostat is applied with a damping parameter of $0.002$, ensuring the rapid dissipation of inertial transients, and a time step of $\Delta t=0.0005$. The systems are equilibrated for $10^7$ integration steps to ensure structural relaxation and remove biases from the initial configurations. Subsequently, simulation data are collected for $10^8$ steps, with coordinates sampled every $10^2$ steps.
\begin{table*}[htb]
\caption{Key parameters and relations}
\label{table_para_rel}
\begin{tabular}{|c|c|c|c|c|}
\hline
\multicolumn{2}{|c|}{\textbf{BRP}} & \multicolumn{2}{c|}{\textbf{LRM}} & \multicolumn{1}{c|}{\textbf{Relation}} \\ \hline
\textbf{Parameter} & \textbf{Notation} & \textbf{Parameter} & \textbf{Notation} &  \\ \hline
\makecell{Number of branches in a single \\ branched ring polymer (BRP)} & $m_{b}$ &  &  & \(m_b=5\) \\ \hline
\makecell{Number of cyclic backbone (CB) \\ i.e Number of BRPs in the system} & $m_c$ & \makecell{Number of ring \\ polymers (RP) in the system} & $m_r$ & \makecell{$m_c = m_r = 9$}\\ \hline
\makecell{Number of branches \\ in the system} & \makecell{$M_b$ \\ ($M_b=m_{b} \times m_c$)} & \makecell{Number of linear \\ polymers (LP) in the system} & $M_l$ & $M_b = M_l$ \\ \hline
\makecell{Number of monomers \\ in a single CB} & $n_c$ & \makecell{Number of monomers \\ in a single RP} & $n_r$ & \makecell{$n_c = n_r = 60$ }\\ \hline
\makecell{Number of monomers \\ in a single branch} & $n_b$ & \makecell{Number of monomers \\ in a single LP} & $n_l$ & \makecell{$n_b = n_l=30$ }\\ \hline
\makecell{Number of monomers \\ in a single BRP} & \makecell{$N_b$ \\ (\makecell{$N_b=n_c+(m_b \times n_b )$ })}  &  &  & $N_b = 210$ \\ \hline
\makecell{Total number of \\ monomers in the system} & \makecell{$N$ \\ ($N = N_b \times m_c )$}  & \makecell{Total number of \\ monomers in the system} & \makecell{$N$  \\ $(N = (m_r \times n_r)$ \\ $+ (n_l \times M_l)$)} & \makecell{$N = 1890$ }\\ \hline
\end{tabular}
\end{table*}
\section{Physical Observables}
We introduce the physical observables to quantify the tracer dynamics from the simulation data. 
Primarily, to characterize the transport behavior of tracer particles in a viscoelastic medium, we computed the mean squared displacement (MSD) of the particles. The ensemble-averaged MSD (eMSD) of $N$ independent trajectories is defined as : 
\begin{equation} \label{EAMSD_eq}
\langle \Delta \boldsymbol{r}(t)^2 \rangle = \frac{1}{N} \sum_{i=1}^{N} |\boldsymbol{r}_i(t) - \boldsymbol{r}_i(0)|^2,
\end{equation}
here, $\boldsymbol{r}(t)$ is the position of the particle at time t, $\boldsymbol{r}(0)$ is the initial reference point of the particle. $\langle ... \rangle$ denotes the ensemble average over $i = 1 \to N$ trajectories.
Moreover, for single-particle analysis, one uses the time-averaged MSD (tMSD) from the individual trajectory, which is defined as \cite{Ralf_JPhysD_2021, Roichman_JCP_2021}:
\begin{equation} \label{Eq:TAMSD}
\overline{\delta r^2(\tau)} = \frac{1}{T_{max} -\tau}{\int_{0}^{T_{max} - \tau} (\bm{r}(t+\tau)-\bm{r}(t))^{2}\, dt},
\end{equation}

where $T_{max}$ is the total length of the trajectory, which is known as the total simulation time, and $\tau$ is the lag time that defines the window size sliding along length of the trajectory. As $\tau$ approaches the long time-scale, this tMSD from a single trajectory becomes inconsistent. To smooth the MSD profile, we have considered the sum over $N$ independent trajectories, ensemble-averaged, time-averaged mean-squared displacement (etMSD), $\left\langle \overline {\delta^2(\tau)} \right\rangle = \frac{1}{N} \sum_{i=1}^{N} \overline {\delta_i^2(\tau)}$. The Brownian diffusion implies that the particle's eMSD equals its individual tMSD over an observed time interval\cite{Burov_Ralf_PCCP_2011}.
For Brownian motion, MSD follows the linear power-law form as a function of time, $\left\langle \overline{\delta r^2(\tau)} \right\rangle \sim 4D_T \tau^\alpha$, where $\alpha$ is known as the local exponent that dictates the diffusion process and $D_T$ is the translational diffusion constant. Numerically, the local exponent is defined as:
\begin{equation}\label{eq:Logarthemic_exponent}
\alpha(\tau) = \frac{\partial \log \left\langle\overline{\delta r^2(\tau)}\right\rangle} {\partial \log \tau}. 
\end{equation}
Here, $\alpha=1$ indicates normal diffusion, and for anomalous diffusion $\alpha\neq1$, i.e., $\alpha < 1$ or $\alpha > 1$ corresponds to subdiffusion or superdiffusion, respectively. The ballistic limit yields $\alpha = 2$. To measure the diffusive behavior, one can define the diffusion constant ($D$) from the MSD relation at the asymptotic limit as: $D_{eff} = \lim\limits_{\tau \to \infty} \left\langle\overline{\delta r^2(\tau)}\right\rangle / 4\tau$ in 2-dimensions.

Particularly, as we consider ABP, whose analytical expression for MSD is given by \cite{Bechinger_Review_2016,Sevilla_PRE_2015},

\begin{equation} \label{Theory: MSD}
\left\langle\overline{\delta r^2(\tau)}\right\rangle 
=4D_T \tau + \frac{2F_a^2 \tau_R}{\gamma^2}\left[\tau + \tau_R\left(e^{-\frac{\tau}{\tau_R}}-1\right)\right],
\end{equation}

where $\tau_R$ is the characteristic time for rotational diffusion. It is related with the rotational diffusion constant ($D_R$), via $\tau_R=\frac{1}{D_R}$. At very short times ($\tau<<\tau_R$), $\left\langle\overline{\delta r^2(\tau)}\right\rangle=4D_T\tau$, the motion is diffusive ($\alpha=1$) with diffusion coefficient $D_T$. At intermediate time ($\tau\approx\tau_R$), $\left\langle\overline{\delta r^2(\tau)}\right\rangle=4D_T\tau + \frac{2F_a^2}{\gamma^2}\tau^2$, the motion is super-diffusive ($\alpha=2$), and at much longer time limit ($\tau>>\tau_R$),$\left\langle\overline{\delta r^2(\tau)}\right\rangle=(4D_T + \frac{2F_a^2\tau_R}{\gamma^2})\tau$, motion becomes diffusive ($\alpha=1$) again associated with enhanced diffusion. The extra term in the diffusion constant arises from active transport.

The non-equivalence of eMSD and tMSD is a manifestation of ergodicity breaking, which is quantified with the so-called ergodicity breaking ($EB$) parameter\cite{Ralf_PRE_2024, Guo_JStatMech_2018}:

\begin{equation} \label{eq: EB Parameter}
EB(\tau) = \frac{\langle (\overline{\delta r^2(\tau)})^2 \rangle}{\langle\overline{\delta r^2(\tau)}\rangle^2} - 1
\end{equation}

Statistically, this parameter is also known as the variance of $\overline{\delta r^2(\tau)}$ over all the trajectories. For ergodic systems, one can expect $EB$ to vanish as $\tau/t \to 0$ \cite{Ralf_NJP_2013, Guo_PRE_2023}, for trajectories longer than all characteristics of the dynamic system. $EB$ parameter depends on both the lag time $(\tau)$ and measurement time $(t)$. For Brownian diffusion, at $\tau << t$, $EB_{BM} = 4\tau / 3t$ \cite{Ralf_NJP_2013, Sokolov_PhyReports_2015,Guo_ChaSolFrac_2023,Guo_PRE_2023, Cherstvy_JPhyA_2014, Ralf_JPhyA_MathTheo_2020, Ralf_JStatMech_2015, Cherstvy_BBA_2016}. For a non-ergodic process, $EB$ gives a finite value, since all individual trajectories produce large variation in tMSD profiles. It is important to note that evaluating the time evolution of $EB(\tau)$ requires significantly more statistics than calculating the MSDs. The definition of the $EB$ (Eq. \ref{eq: EB Parameter} )in terms of variation is written as \cite{Ralf_JPhysD_2021,Ralf_PRE_2024,Ralf_NJP_2013, Ralf_JStatMech_2015, Cherstvy_BBA_2016},

\begin{equation} \label{eq: EBP varibale}
EB(\tau) = \langle \xi^2(\tau) \rangle - 1,
\end{equation}
Here, the variable $\xi(\tau) = \frac{\overline{\delta r^2(\tau)}}{\langle\overline{\delta r^2(\tau)}\rangle}$ quantifies relative deviations of individual tMSD realizations around their mean divided by etMSD. The system is ergodic when $\xi(\tau)$ converges to unity for all individual trajectories at $\tau/t << 1$, indicating the complete reproducibility of all tMSD trajectories within the ensemble \cite{Ralf_NJP_2013, Ralf_PRE_2024}. 

Typically, anomalous diffusion yields non-Gaussian dynamics. For a $d$-dimensional system, the degree of non-Gaussianity can be examined using the non-Gaussian parameter \cite{Jeon_MacMol_2022, Ralf_PRE_2024, Ghosh_NJP_2016, Cherstvy_BBA_2016,BoWang_PNAS_2009},

\begin{equation} \label{eq: NGP}
G(\tau) = \frac{d}{d+2}\frac{\langle\overline{\delta r^4(\tau)}\rangle }{\langle\overline{\delta r^2(\tau)}\rangle^2}-1,
\end{equation}

where the fourth moment $\overline{\delta r^4(\tau)} $ serves as a magnifying parameter that enhances the outliers in the displacement data. A perfect Gaussian process implies $G(\tau) = 0$. The deviation from Gaussian behavior yields $G(\tau) < 0$ for narrow distribution, and $G(\tau) > 0$ corresponds to broader distribution.

To understand the microscopic displacement statistics probabilistically and identify transient active trapping in polymeric environments, we compute the self-part of the van-Hove correlation function (VHCF). For an isotropic medium, the behavior along all directions should be identical. The VHCF is mathematically defined as \cite{Zhao_PCCP_2022,simma2025competing},
\begin{equation} \label{eq: vanHove}
P(\Delta r; \tau) \equiv \langle \delta (\Delta r - (r(t+\tau) - r(t)))\rangle,
\end{equation}
where $\Delta r$ is the spatial displacement over a specific lag time $\tau$. The VHCF is Gaussian for Brownian diffusion, and it shows non-Gaussian behavior at intermediate times for non-equilibrium active particles.

In order to investigate the interaction of polymers with the tracer particle locally, we have computed the radial distribution function, $g(r)$, which is defined as \cite{SPSingh_PRE_2020}:
\begin{equation} \label{eq: RDF}
g(r) = \frac{N(r)}{2\pi r dr \rho N_p},
\end{equation}
Here, $N(r)$ is the average number of monomers around the tracer in a circular shell of radius r and thickness $dr$, $2\pi r dr$ represents the area of that shell, and $\rho$ is the average number density of monomers in the system.

The conformations of the polymers affect the dynamics of the tracer particle. The most conventional way to understand the size of a polymer is the radius of gyration \cite{Rudnick_Science_1987, Kotkar_JPhyChemB_2024}, 

\begin{equation} \label{eq: RG}
\langle R_g^2\rangle = \frac{1}{S} \sum_{i=1}^S (\boldsymbol{r}_i - \boldsymbol{r}_{cm})^2,
\end{equation}

where $\boldsymbol{r}_i$ and $\boldsymbol{r}_{cm}$ are the positions of $i^{th}$ monomer, and polymer center of mass respectively. $S$ is the number of monomers in each polymer. The radius of gyration represents the average spread of a random walk. To understand the structural arrangements of polymers, we compute the shape descriptor, known as the asphericity parameter \cite{ErwinFrey_PRL_2007},

\begin{equation} \label{eq: Ashericity}
A = \frac{(\lambda_1-\lambda_2)^2}{(\lambda_1+\lambda_2)^2}
\end{equation}
where $\lambda_1$ and $\lambda_2$ are the eigenvalues of the gyration tensor \cite{Abel_SoftMatter_2021, Kotkar_JPhyChemB_2024}.  $\langle A \rangle$ takes values from $0$ to $1$, where $0$ indicates a circular configuration, and $1$ corresponds to complete stretched conformations.

An active tracer diffuses through space over time. More general spatiotemporal information is encrypted in the self-intermediate scattering function (ISF), \cite{Franosch_SciRep_2016} 
\begin{equation}\label{eq: ISF}
F_s(q,\tau)=\langle e^{-i\boldsymbol{q}\cdot[\boldsymbol{r}(\tau)-\boldsymbol{r}(0)]} \rangle
\end{equation}

where $r(\tau)$ and $r(0)$ are the positions of the particle at time $\tau$ and 0 respectively. $q$ is the characteristic function of displacement in Fourier space and $\langle...\rangle$ is the average over various independent trajectories. This function describes the motion of the particle at a lag time $\tau$ on a length scale $2\pi/q $. It is the Fourier transform of the position probability density of the particle. Thus all the lower-order moments of the positions of the active particle can be acquired by taking derivatives of the ISF with respect to the wavevector \cite{Franosch_SciRep_2016,Franosch_PRE_2024}. This quantity can be measured directly in scattering experiments \cite{Berne_BPJ_1977} such as dynamic light scattering. It can also be obtained from the statistical information collected in single-particle tracking.

\section{Results}
\label{sec-results}

\subsection{Effect of Stiffness: Fixed Packing Fraction}
We use LRM as a baseline model and study tracer dynamics in a BRP environment formed by ring polymers serving as the backbone and linear polymers acting as branches. To enable direct comparison, we present the results for the two systems under identical conditions. The packing fraction is fixed at $\phi = 0.1$ for both systems. Any observed differences can therefore be primarily attributed to the change in the polymer topology.

\begin{figure}
    \includegraphics[width=8cm]{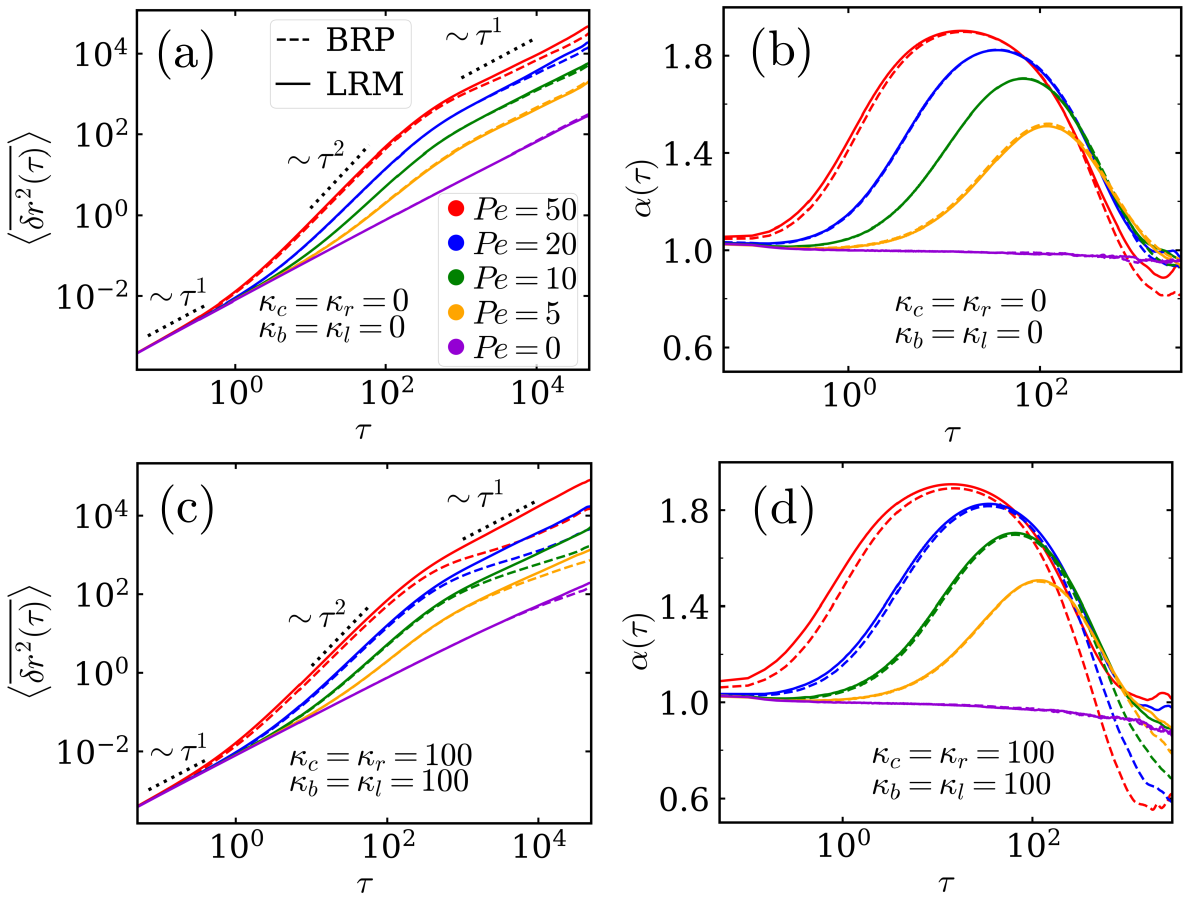}
  \caption{Log-log plot of the ensemble-averaged time-averaged mean squared displacement (etMSD), $\langle\overline{\delta r^2(\tau)}\rangle$, as a function of lag time $\tau$ for an active tracer at varying P\'eclet numbers ($Pe$) in the Branched Ring Polymer (BRP, dashed lines) and Linear-Ring Mixture (LRM, solid lines) systems. Results are shown for (a) fully flexible ($\kappa = 0$) and (c) stiff ($\kappa = 100$) polymers, with the corresponding local exponents, $\alpha(\tau)$, plotted on a log-linear scale in (b) and (d), respectively. The black dotted lines indicate the respective reference power-law scalings ($\sim \tau^1$ and $\sim \tau^2$).}
  \label{taMSD_Beta_diff_Pec}
\end{figure}

\begin{figure}
\centering
\includegraphics[width=8.0cm, height=6cm]{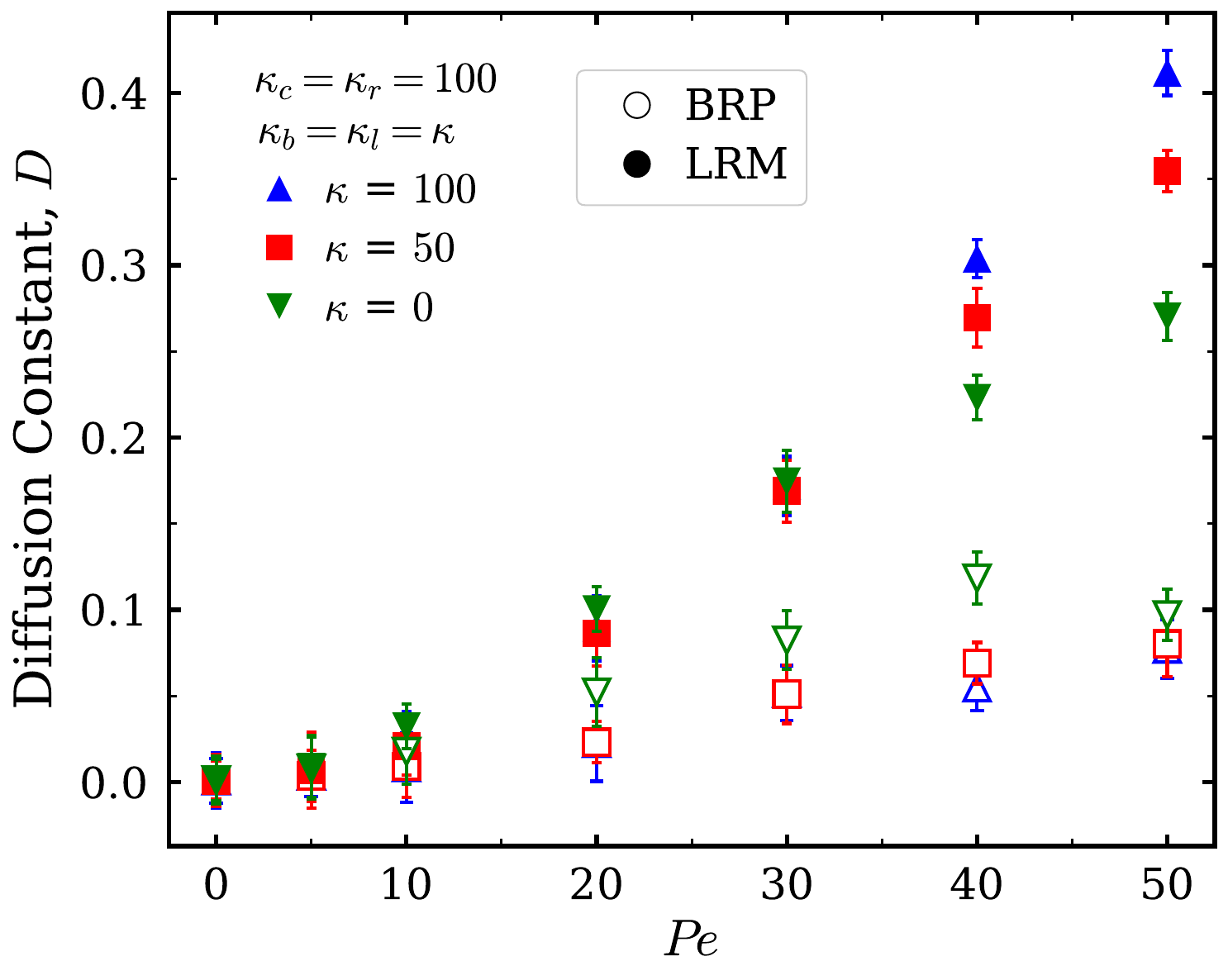}
\caption{Diffusion constant, $D$, of the active tracer as a function of propulsion strength, $Pe$, in the BRP (open markers) and reference LRM (filled markers) environments. While the cyclic backbones and ring polymers are kept stiff ($\kappa_c = \kappa_r = 100$), the bending stiffness of the branches and linear chains ($\kappa_b = \kappa_l = \kappa$) is systematically varied from flexible ($\kappa = 0$, green) to semiflexible ($\kappa = 50$, red) and stiff ($\kappa = 100$, blue). Across all measured activities, the tracer consistently exhibits slower diffusion in the branched topology, with increasing branch stiffness suppressing $D$ in BRP but getting enhanced in LRM systems.}
\label{D_Pe}
\end{figure}
\begin{figure}
\centering
\includegraphics[width=8cm]{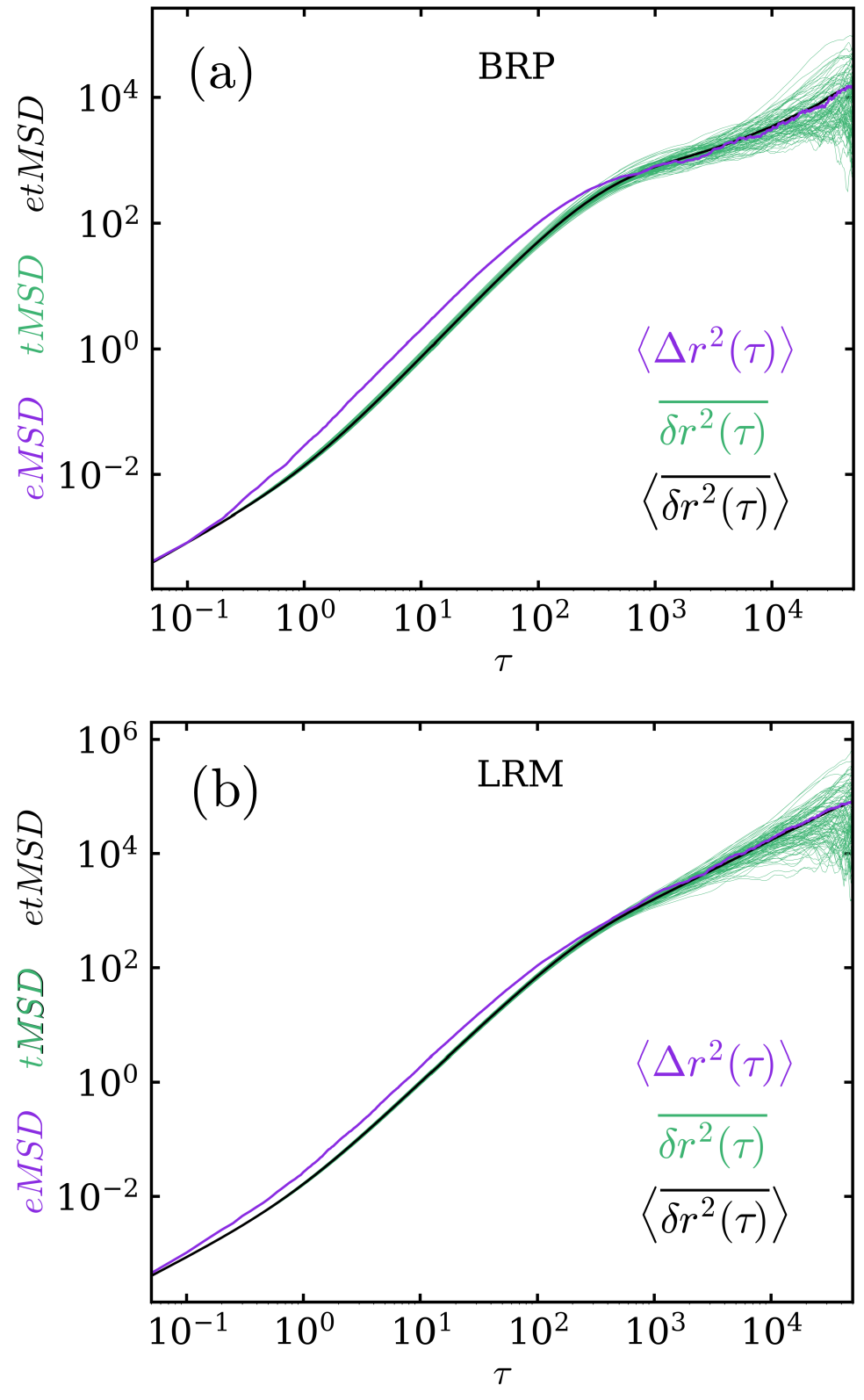}
\caption{Comparison of the eMSD $\langle\Delta r^2(\tau)\rangle$ (violet), individual tMSDs $\overline{\delta r^2(\tau)}$ (green), and etMSD $\langle\overline{\delta r^2(\tau)}\rangle$ (black) for a highly active tracer ($Pe = 50$) in stiff polymeric environments  ($\kappa = 100$) : (a) the BRP system ($\kappa_c = \kappa_b = 100$) and (b) the reference LRM system ($\kappa_r = \kappa_l = 100$). The intermediate-time discrepancy between the eMSD and etMSD reflects transient weak ergodicity breaking.} 
\label{All_MSD}
\end{figure}
\subsubsection{Time-averaged MSD, Logarithmic exponent, and, Diffusion Constant}

To understand the role of polymer architecture independent of purely steric crowding effects, we systematically varied the propulsion strength ($Pe$) of ABP from $0$ to $50$ and the bending stiffness ($\kappa$) of the polymer constituents.

We analyze etMSD (Eq.\ref{Eq:TAMSD}), $\left\langle\overline{\delta r^2(\tau)}\right\rangle$, and its logarithmic exponent (Eq.~\ref{eq:Logarthemic_exponent}), $\alpha(\tau)$, to quantitatively characterize the dynamic regimes; see Fig.~\ref{taMSD_Beta_diff_Pec}. In the polymeric systems where all the polymer constituents are fully flexible ($\kappa = 0$), passive tracers ($Pe = 0$) exhibit conventional Brownian diffusion ($\alpha \approx 1$) at all time regimes (Fig.~\ref{taMSD_Beta_diff_Pec}(a,b) and Movie S1, Movie S2). Upon introduction of activity ($Pe$), we observe the canonical sequence of active transport: short-time normal diffusion, intermediate self-propelled superdiffusion ($\alpha > 1$), and a return to long-time normal diffusion. In this flexible limit, the BRP system exhibits nearly identical scaling to the reference LRM system, with the only notable divergence at long timescales for the highest activity ($Pe \approx 50$) with $\alpha < 1$ (Movie S3 \& Movie S4). 

The distinct dynamical properties begin to emerge in LRM topologies in the stiff limit ($\kappa = 100$); see Fig.~\ref{taMSD_Beta_diff_Pec}(c,d). We note that in this limit, the crowding effect is severe enough that even a passive tracer ($Pe=0$) experiences significant long-time subdiffusion in both BRP and LRM melts (Movie S5 \& Movie S6). However, introducing activity endows the tracer with persistent motion, enabling it to traverse extended length scales and effectively probe the detailed topological distribution. While the active tracer in the LRM system maintains its standard superdiffusive-to-diffusive sequence, the tracer in the stiff BRP matrix strongly deviates, plunging into a pronounced subdiffusive regime ($\alpha < 1$) at large $\tau$. Counterintuitively, this long-time anomalous subdiffusion in the BRP system becomes monotonically more pronounced with increasing $Pe$. Representative time-resolved trajectories (color-coded) of a highly active tracer ($Pe=50$) across the different polymer conformations formed by varying the component stiffnesses ($\kappa$) are shown in Figs.~\ref{BRP_Conf_Trj} and ~\ref{LRM_Conf_Trj} for the BRP and LRM systems, respectively, clearly illustrating localized trapping in the stiff BRP matrix compared to extended exploration in the LRM. To further resolve how the tracer dynamics depend on the individual component stiffnesses, the etMSD of the active tracer ($Pe=50$) for different stiffness combinations of the structural components is shown in Fig.~\ref{tamsd_diff_Kbr_Kr}.

Furthermore, Fig.~\ref{taMSD_Beta_diff_Pec}(b, d) reveal a distinct kinetic feature: highly active particles transition out of the intermediate superdiffusive regime earlier than their less active counterparts. This faster crossover is driven by the frequent collisions between the highly active tracer and the polymer matrix. At higher $Pe$, frequent steric interactions rapidly decorrelate the active particle's orientation, effectively reducing its rotational relaxation time, $\tau_R$. Consequently, the tracer's motion randomizes faster. In the unlinked LRM system, this rapid randomization directly leads to an earlier onset of enhanced diffusion, $D_{\text{eff}} = D_T + (F_a^2 / 2\gamma^2)\tau_R$. In the stiff BRP system, however, this diffusive limit is preempted by severe topological caging. The stiffness dependence on $\alpha$ further distinguishes the two systems, revealing the persistence of and departure from normal diffusion (Fig.~\ref{beta_diff_Kbr_Kr}). 

This distinct localization arises directly from the conformational constraints inherent to the branched architecture. Increasing the stiffness in the BRP system enforces the interconnected stretched cyclic backbones and elongated branches to create extended, star-like conformations; see Fig.~\ref{BRP_Conf_Trj}. These rigid structures form mesh-like topological cages. Due to its self-propulsion, an active particle rapidly penetrates these localized pockets but occasionally gets trapped, leading to prolonged localization events (Movie S7). Conversely, the reference LRM system where linear polymers are not directly connected to the rings lacks these interconnected constraints, yielding a relatively homogeneous crowded distribution even in the stiff limit (Fig.~\ref{LRM_Conf_Trj} \& Movie S8). This establishes that the observed arrested dynamics are fundamentally a manifestation of the branched topology.

To quantify the macroscopic transport behavior, we calculated the long-time diffusion constant, $D$, as a function of the propulsion strength, $Pe$ (Fig.~\ref{D_Pe}). Naturally, $D$ increases monotonically with activity across biological systems \cite{Mino_PRL_2011,Howse_PRL_2007}. How the surrounding polymeric environment affects active dynamics depends critically on the interplay between polymer stiffness and its topology \cite{Zhao_PCCP_2022}. 
As shown in Fig.~\ref{D_Pe}, the tracer consistently exhibits slower diffusion in the branched topology compared to the reference LRM system across all measured activities and rigidities, including the fully flexible limit. In the low-activity regime ($Pe \leq 10$), the macroscopic diffusion is largely insensitive to the bending rigidity ($\kappa$).  

Conversely, a significant deviation between the two topologies emerges at higher activities. In the unlinked LRM environment, increasing the stiffness of the linear chains actually facilitates the tracer diffusion. However, in the BRP matrix, increasing branch stiffness severely slows down particle mobility (Fig.~\ref{D_Pe}, blue marker).  This opposing effect of the polymer matrix perfectly demonstrates the counterintuitive uniqueness of the BRP environment: while a stiff, unlinked system only alters the homogeneous crowding distribution to aid active transport, a stiff, highly branched network creates topological cages that drive the system toward slower dynamics.

\subsubsection{Transient Ergodicity Breaking}

\begin{figure*}
\centering
\includegraphics[width=16cm]{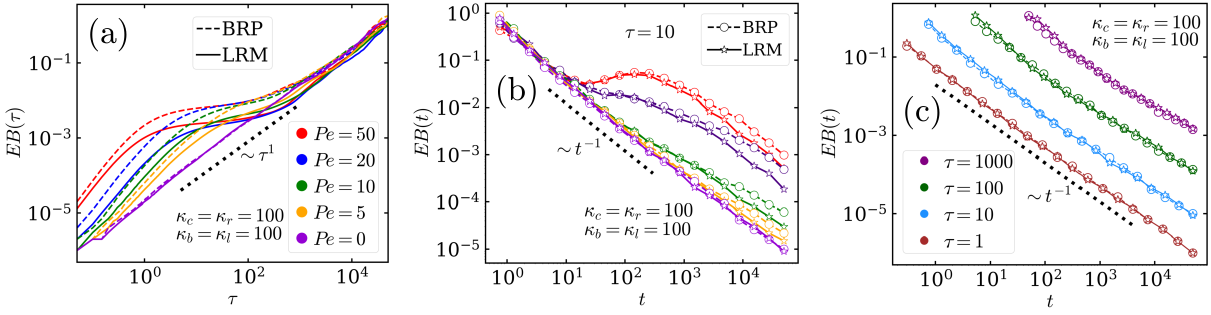}
\caption{Ergodicity breaking parameter, $EB$, in the stiff polymer limit ($\kappa = 100$) for the BRP (dashed lines) and LRM (solid lines) systems, as a function of (a) lag time ($\tau$) and (b) measurement time ($t$) at a fixed lag time $\tau = 10$ for varying propulsion strengths ($Pe = 0, 5, 10, 20, 50$). (c) $EB(t)$ versus measurement time across different lag times ($\tau = 1, 10, 100, 1000$) for a passive tracer ($Pe = 0$). The thick black dotted lines indicate the asymptotic scaling $EB \sim \tau^1$ in (a) and $EB \sim t^{-1}$ in (b, c). The color scheme in (a) and (b) follows Fig.~\ref{taMSD_Beta_diff_Pec}.}
\label{EB_Parameter}
\end{figure*}

The anomalous sub-diffusion, as shown in Fig. \ref{taMSD_Beta_diff_Pec}(c) results the heterogeneity in the tracer dynamics, which triggers the non-ergodic behavior. To further elucidate the nature of weak ergodicity breaking (WEB) \cite{Barkai_PRL_2005, Bouchaud_JPhysIFrance_1992, Ralf_PCCP_2014, Ralf_JStatMech_2015, Fulinski_PRE_2014}, a characteristic of heterogeneous crowded systems with caging even without attractive centers\cite{Ralf_PRE_2024, Ralf_NJP_2013, Guo_ChaSolFrac_2023, Cherstvy_JPhyA_2014}, we compared three quantities such as eMSD, tMSD, and etMSD as shown in Fig. \ref{All_MSD}. In an ergodic system, the eMSD ($\langle\Delta r^2(\tau)\rangle$), matches with the etMSD ($\langle\overline{\delta r^2(\tau)}\rangle$) \cite{Sokolov_PhyReports_2015, Ralf_JStatMech_2015, Ralf_JPhysD_2021, Cherstvy_BBA_2016, Fulinski_PRE_2014}. Fig.~\ref{All_MSD} compares $\langle\Delta r^2(\tau)\rangle$ and  $\langle\overline{\delta r^2(\tau)}\rangle$ in the stiff limit ($\kappa=100$) under high activity ($Pe \approx 50$) for both the BRP (top) and LRM (bottom) systems. We observe a transient mismatch between $\langle\Delta r^2(\tau)\rangle$ and $\langle\overline{\delta r^2(\tau)}\rangle$ at the intermediate timescale in the stiff BRP (Fig.~\ref{All_MSD} (a)) and LRM (Fig.~\ref{All_MSD} (b)). This divergence indicates that individual trajectories are moderately sparse, with varying residence times inside the topological cages.

Notably, in the fully flexible limit ($\kappa = 0$), the transient discrepancy within the BRP system is even more prominent than in the stiff limit for a highly active particle ($Pe = 50$); see Fig.~\ref{all_MSD_K_0}. Physically, while the stiff BRP network creates a few isolated, large localized pockets, the flexible BRP matrix forms numerous tiny, dynamic pockets, often not so well defined. These dynamic constraints force frequent pocket hopping, complemented by short diffusion events for the tracer, enhancing the intermediate-time divergence between $\langle\Delta r^2(\tau)\rangle$ and $\langle\overline{\delta r^2(\tau)}\rangle$. Regardless of matrix flexibility, the BRP architecture consistently shows a more prominent intermediate-time deviation than the equivalent LRM system, which we subsequently clarify using the ergodicity-breaking parameter. 

\subsubsection{Ergodicity Breaking Parameter}

To quantitatively capture the onset of WEB and the trajectory-to-trajectory variation, we compute the ergodicity-breaking parameter, $EB(\tau)$, see Eq.~\ref{eq: EB Parameter}. In ergodic processes, individual tMSD trajectories converge to the eMSD, resulting in $EB \to 0$ in the asymptotic limit \cite{Sokolov_PhyReports_2015, Ralf_JStatMech_2015, Ralf_JPhysD_2021, Cherstvy_BBA_2016, Fulinski_PRE_2014, Guo_ChaSolFrac_2023}. For a free ABP, $EB$ exhibits a linear temporal variation across the entire range of activities considered; see Fig.~\ref{EBP_freespace}(a). Fig \ref{EB_Parameter}(a) illustrates the variation of $EB(\tau)$ with lag time $\tau$ in the stiff polymeric environments ($\kappa_c=\kappa_r=\kappa_b=\kappa_l=100$). For a passive tracer ($Pe = 0$), $EB$ increases linearly with lag time ($EB \sim \tau^1$) \cite{Ralf_JPhyA_MathTheo_2020}. Finite activity ($Pe > 0$) consistently elevated the $EB(\tau)$ parameter, reflecting increasingly non-ergodic behavior at higher $Pe$. A distinct plateau ($EB \sim \tau^0$) is observed at the intermediate lag time scale \cite{Guo_ChaSolFrac_2023, Guo_PRE_2023, Ralf_JPhyA_MathTheo_2020}, implying a constant variance in the spreading of individual tMSD trajectories around the mean. At the logarithmic time scale, all the $EB(\tau)$ curves converge to the asymptotic linear scaling behavior across all $Pe$. Most importantly, in general, the branched architecture (BRP) systematically exhibits higher $EB$ values than the reference mixture (LRM). This confirms that the interconnected BRP topology resulted in a stronger ergodicity breaking by imposing stronger spatial heterogeneity and restricting diffusion. 

Further, one can examine ergodicity breaking with respect to the measurement time, $t$ \cite{Maggi_PCCP_2026, Barkai_PRE_2009}. Fig \ref{EB_Parameter}(b) represents the evolution of $EB(t)$ as a function of $t$ for a fixed lag time $\tau = 10$ in the same stiff limit. For the passive tracer, $EB(t)$ strictly follows the expected scaling characteristic ($\sim t^{-1}$) of standard Brownian motion. This is further verified in Fig. \ref{EB_Parameter}(c), which confirms that for a passive tracer ($Pe=0$), $EB(t)$ maintains the $t^{-1}$ decay across varying lag times ($\tau$) \cite{Ralf_PRE_2010}. Irrespective of its activities in free space, $EB$ displays an inverse dependence on measurement time ($t$) at a fixed lag time,$\tau=10$ (Fig.~\ref{EBP_freespace}(b)). The correlation between the coordinate in a trajectory decreases as we increase $\tau$, resulting in a corresponding upward shift in the magnitude of $EB(t)$ \cite{Maggi_PCCP_2026, Ralf_PRE_2010}. As illustrated in Fig.~\ref{EB_Parameter}(b), the magnitude of $EB(t)$ systematically increases with $Pe$, with the BRP system consistently exhibiting higher $EB(t)$ values than the equivalent LRM system across all measured activities. These ergodicity-breaking trends persist even in the fully flexible limit ($\kappa=0$), as detailed in Fig \ref{EB_time_Diff_Pe}. 


\subsubsection{van-Hove correlation function}

\begin{figure*}
\includegraphics[width=12cm]{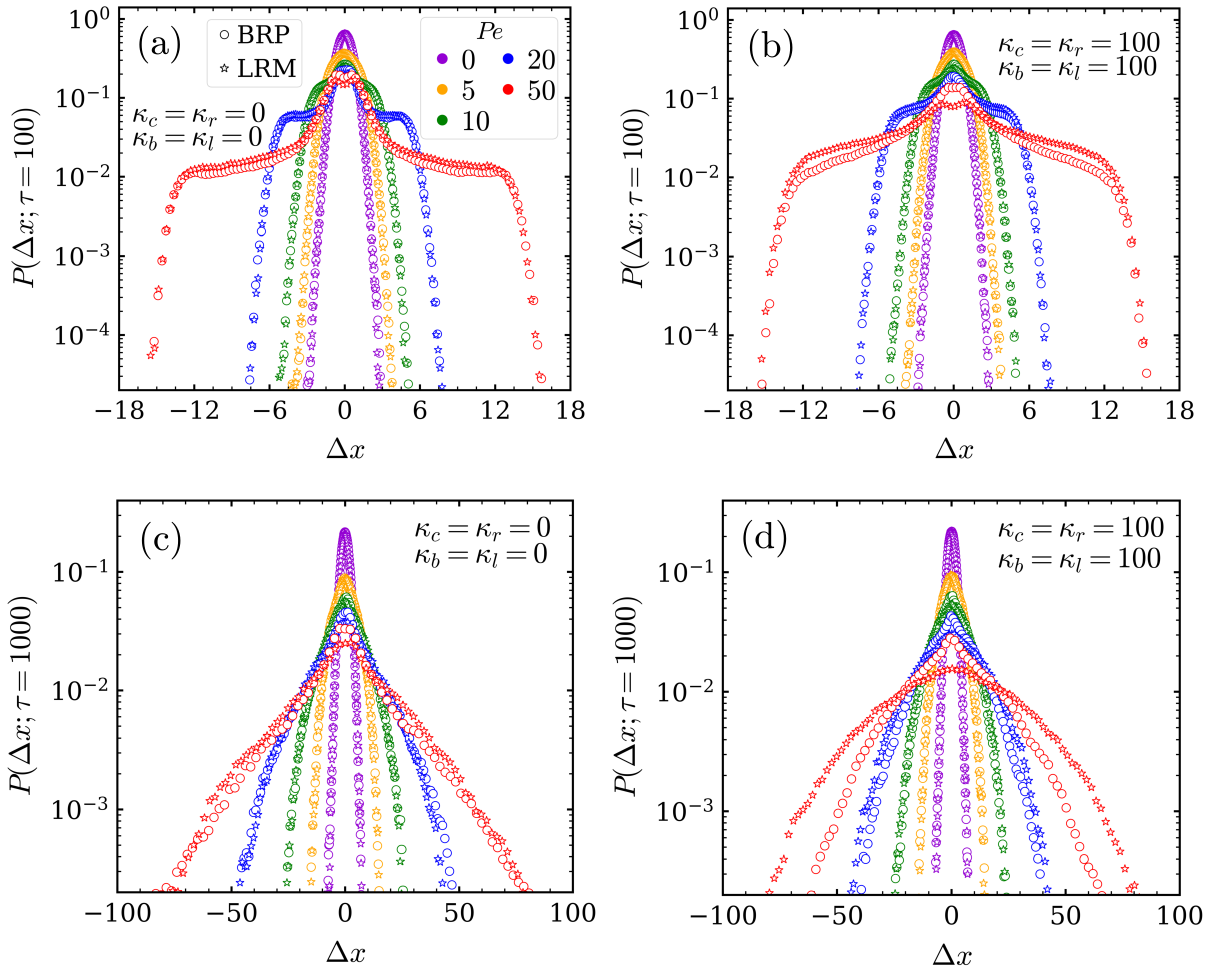}
\caption{Self-part of the van Hove correlation function, $P(\Delta x; \tau)$, of the ABP in the BRP (circles) and LRM (stars) systems, evaluated at lag times $\tau = 100$ (top row: (a), (b)) and $\tau = 1000$ (bottom row: (c), (d)), for fully flexible ($\kappa = 0$, left column: (a), (c)) and stiff ($\kappa = 100$, right column: (b), (d)) polymers. Colors denote different P\'eclet numbers ($Pe = 0, 5, 10, 20, 50$) as indicated in (a).}
\label{van_Hove}   
\end{figure*}

To probabilistically characterize the constrained dynamics induced by polymeric environments with distinct architectures, we compute the van-Hove correlation function (VHCF), $P(\Delta x; \tau)$ (Eq.~\ref{eq: vanHove}). Fig.~\ref{van_Hove} compares the VHCF in the flexible ($\kappa = 0$) and stiff ($\kappa = 100$) limits for the BRP (circles) and LRM (stars) systems. The passive tracer ($Pe = 0$) exhibits nearly Gaussian profiles in both flexible and stiff polymeric media at lag time $\tau = 100$ (Fig.~\ref{van_Hove}(a, b)), as passive diffusion relies on thermal fluctuations that produce small displacements at shorter lag times. However, active tracers ($Pe > 0$) exhibit pronounced non-Gaussian profiles. In particular, a prominent central peak emerges at high propulsion strength ($Pe = 50$) in the flexible medium, whereas its central peak intensity is reduced in the stiff polymeric medium.

Fig.~\ref{van_Hove}(c) shows the variation of VHCF at a longer lag time, $\tau = 10^3$, for a fully flexible medium ($\kappa = 0$). Consistent with its normal diffusive MSD, the passive tracer maintains a Gaussian distribution across both flexibility limits (Fig.~\ref{van_Hove}(c, d)). Increasing activity broadens the VHCF and drives a departure from Gaussianity ($Pe > 0$). Strong non-Gaussian profiles with a sharp central peak are observed at high P\'eclet number ($Pe \approx 50$). The fully flexible BRP and LRM media create locally confined domains even at dilute polymer packing. These small, flexible cages obstruct tracer dynamics; highly active tracers encounter these cages more frequently, resulting in a pronounced peak intensity; as a result, we noticed the proper non-Gaussian profile (Fig.~\ref{van_Hove}(c)).

Subsequently, Fig.~\ref{van_Hove}(d)  illustrates the VHCF in a stiff ($\kappa = 100$) polymeric medium at $\tau = 1000$. While moderate activity leads to non-Gaussian trends with slight differences between the two topologies. A stark distinction between the BRP and LRM, VHCF profiles is explicitly established at higher propulsion strength ($Pe = 50$). The stiff BRP medium forms interconnected traps that profoundly restrict tracer dynamics. However, the stiff LRM medium cannot form persistent cages because the unlinked linear and ring polymers are well separated and diffuse independently. As a result, we observe a nearly Gaussian profile for $Pe = 50$ in the stiff LRM medium, whereas the VHCF in the BRP matrix shows a strongly non-Gaussian profile due to topological caging. This clearly demonstrates that polymer architecture governs microscopic tracer dynamics even at low packing fractions.

\begin{figure*}
\centering
\includegraphics[width=15cm]{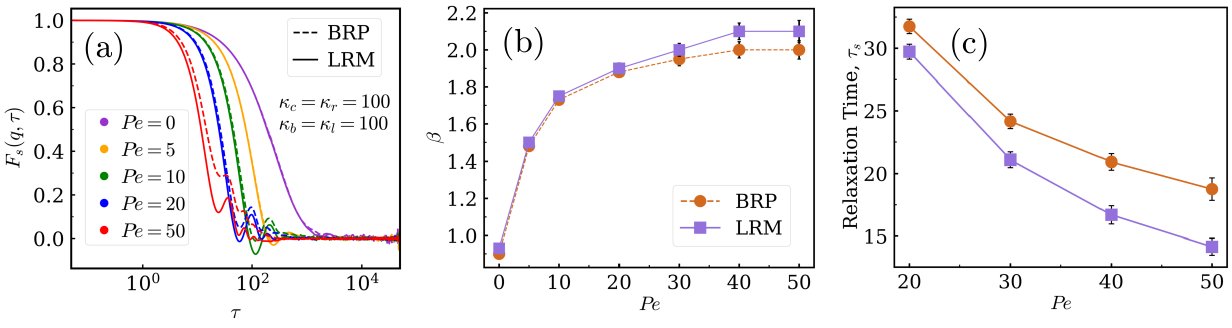}
\caption{(a) Time evolution of the self-intermediate scattering function, $F_s(q,\tau)$, evaluated at wave vector $q = 1$ for varying propulsion strengths ($Pe$) in the stiff BRP (dashed lines) and LRM (solid lines) systems ($\kappa_c = \kappa_b = \kappa_r = \kappa_l = 100$). Dependence of the corresponding (b) compressed-exponential stretching exponent, $\beta$, and (c) structural relaxation time, $\tau_s$, on $Pe$ obtained from fitting eq.~\ref{eq:ISF_active}.}
\label{ISF_Beta_tauS}
\end{figure*}
\subsubsection{Self-intermediate scattering function $(F_{s}(q,\tau))$ and relaxation dynamics of the tracer}

To characterize the spatiotemporal relaxation and length-scale dependence of the tracer, we compute the self-intermediate scattering function (ISF), $F_{s}(q,\tau)$ (Eq.~\ref{eq: ISF}). For a passive tracer following Brownian dynamics, the ISF exhibits a pure exponential decay,

\begin{equation}
F_{s}(q,\tau) = \exp(-D q^2 \tau) = \exp\left(-q^2\left[\frac{\tau}{\tau_s}\right]\right),
\label{eq:ISF_Brownian}
\end{equation}

where the diffusion constant $D$ is defined via the relaxation time $\tau_s$ ($D = 1/\tau_s$).

Fig.~\ref{ISF_Beta_tauS}(a) illustrates the time evolution of $F_{s}(q,\tau)$ at a fixed wave vector $q=1.0$ in the stiff limit ($\kappa_c=\kappa_b=100$, $\kappa_r=\kappa_l=100$) across varying $Pe$. Introducing activity systematically accelerates decay in both topologies due to enhanced active propulsion. At intermediate timescales, persistent motion induces oscillations in $F_{s}(q,\tau)$ whose amplitudes grow with $Pe$, marking the active ballistic regime \cite{Franosch_SciRep_2016}. 

While both systems share this generic decay profile, the BRP matrix exhibits slower overall relaxation. At high activity, the ISF decay in BRP is distinctly stalled compared to the unlinked LRM baseline, reflecting localized dynamics driven by topological caging. Due to this persistent motion, the ISF deviates from the traditional exponential form (Eq.~\ref{eq:ISF_Brownian}) and follows a compressed exponential scaling:

\begin{equation}
F_{s}(q,\tau) = \exp\left(-\left[\frac{\tau}{\tau_s(q)}\right]^\beta\right),
\label{eq:ISF_active}
\end{equation}

where $\beta$ is the stretching exponent \cite{ramakrishnan1987, Kohlrauch_1854, Watts_Williams_1970}. We fit Eq.~\ref{eq:ISF_active} to our data using $\beta$ and $\tau_s$ as fitting parameters to quantify the relaxation of the tracer at different activity strengths (see Fig.~\ref{ISF_Fit}).

As $Pe$ increases, $\beta$ rises monotonically ($\beta > 1$), capturing the ballistic escape from transient cages (Fig.~\ref{ISF_Beta_tauS}(b)). At high $Pe$, $\beta$ plateaus at a slightly lower value in BRP ($\approx 2.0$) than in LRM ($\approx 2.1$), as frequent caging interruptions curb the tracer's persistent runs.

Defining the relaxation time $\tau_s$ as the time where $F_s(q,\tau_s) = 1/e$, Fig.~\ref{ISF_Beta_tauS}(c) shows that $\tau_s$ decreases monotonically with $Pe$ in both systems due to faster active exploration. However, $\tau_s$ remains systematically longer in the BRP matrix than in the LRM system, with the gap widening at higher $Pe$. This aligns with the etMSD trends (Fig. \ref{taMSD_Beta_diff_Pec}): highly active tracers possess longer persistence lengths, causing them to encounter and become trapped in branch-formed pockets more frequently in BRP environments. Escape takes longer, delaying the decay of density fluctuations relative to the unlinked LRM system.

Finally, examining spatial length scales via Fig.~\ref{ISF_Kwv_1_10} ($q=1$ vs. $q=10$) reveals that increasing the wave vector accelerates temporal ISF decay across all propulsion strengths. At these reduced length scales ($q=10$), matrix topology exerts a nominal effect on spatial self-correlations, yielding overlapping BRP and LRM decay profiles for $Pe \le 10$. This length-scale segregation confirms that extended branched architectures dictate macroscopic transport via topological caging at larger length scales ($q=1$), whereas short-range active diffusion remains governed by local particle--polymer interactions.

\subsubsection{Non-Gaussian parameter, \texorpdfstring{$G(\tau)$}{G(tau)} }

\begin{figure}
    \includegraphics[width=9cm]{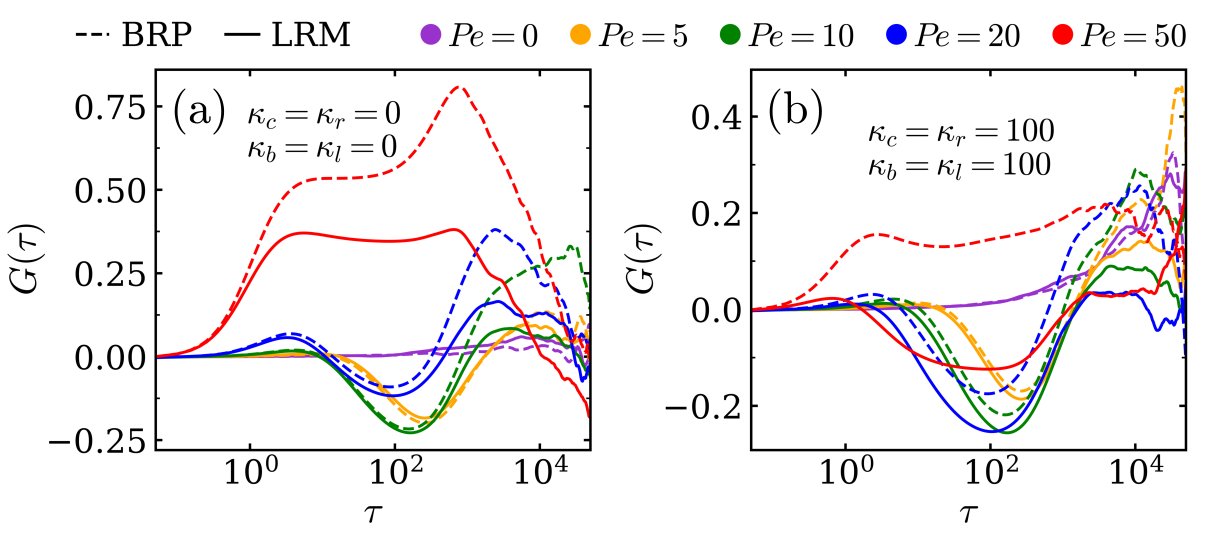}
  \caption{Log-linear plot of the time evolution of the non-Gaussian parameter, $G(\tau)$, for an ABP at varying propulsion strengths ($Pe = 0, 5, 10, 20, 50$) in the BRP (dashed lines) and reference LRM (solid lines) systems, composed of (a) fully flexible ($\kappa = 0$) and (b) stiff ($\kappa = 100$) polymers.}
  \label{NGP_diff_Pec}   
\end{figure}

To quantitatively analyze deviations from Gaussian displacement statistics, we compute the non-Gaussian parameter (NGP), $G(\tau)$ (Eq.~\ref{eq: NGP}), which corresponds to one-third of the excess kurtosis \cite{Zheng_HartmutLowen_PhyRevE_2013, BARAL_Chaos_Solutions_Fractals_2024, Baouche_NJP_2024, Hahn_Jingyu_PhysRevE_2020, Vorselaars_PhysRevE_2007, Nakai_Takashi_PhysRevE_2020, ShotaGoto_Kim_JCP_2021}. Standard Brownian motion yields $G(\tau) = 0$, whereas purely ballistic motion in two dimensions approaches approximately $ G(\tau) \approx -0.5$ \cite{Loewe_SoftMatter_2024}. Specifically, positive values ($G(\tau) > 0$) reflect displacement distributions with narrow central peaks and broad tails (leptokurtic), whereas negative values ($G(\tau) < 0$) indicate flatter peaks and narrow tails (platykurtic) \cite{BARAL_Chaos_Solutions_Fractals_2024, Chen_Macromolecules_2024}.

Fig.~\ref{NGP_diff_Pec} illustrates the time evolution of $G(\tau)$ for distinct P\'eclet numbers ($Pe = 0, 5, 10, 20, 50$) in the BRP (dashed lines) and LRM (solid lines) systems. We first consider a fully flexible matrix ($\kappa_b = \kappa_c = \kappa_l = \kappa_r = 0$, Fig.~\ref{NGP_diff_Pec}(a)). The passive tracer ($Pe = 0$) exhibits nearly Gaussian behavior, with $G(\tau) \approx 0$ (violet line) in both media, as the dilute polymeric environment ($\phi = 0.1$) imposes minimal steric trapping to the passive tracer. However, active tracers ($Pe > 0$) exhibit platykurtic non-Gaussian behavior at intermediate times ($G(\tau) \approx -0.2$ for $Pe = 5, 10, 20$), followed by a crossover to leptokurtic behavior ($G(\tau) > 0$) for $Pe = 10$ and $20$ in the long-time limit ($\tau \to \infty$). This crossover occurs because persistent directed motion dominates at intermediate timescales, whereas at long times the polymeric medium obstructs tracer dynamics by creating local cages. This caging effect is more pronounced in the BRP matrix, contributing to systematically higher $G(\tau)$ values than in the LRM system. At high activity ($Pe = 50$), $G(\tau)$ remains strictly positive across all observed timescales due to strong dynamical heterogeneity. The highly active tracer rapidly encounters and is restricted within flexible polymer pockets, an effect that is significantly stronger in the interconnected BRP medium than in the unlinked LRM baseline systems (Fig.~\ref{NGP_diff_Pec}(a), red line).

In the stiff ($\kappa_b = \kappa_c = \kappa_l = \kappa_r = 100$) polymeric environment (Fig.~\ref{NGP_diff_Pec}(b)), the passive tracer ($Pe = 0$) shows positive non-Gaussian deviations at long times in both systems due to steric hindrance from the extended structure of the rigid polymers. For moderate activities ($Pe \leq 20$), the active tracer shows negative $G(\tau)$ values (reaching $G(\tau) \approx -0.25$), reflecting pronounced persistent motion at intermediate times, followed by occasional caging at long times. Moreover, a highly active tracer ($Pe = 50$) exhibits a stark topological contrast between the BRP and LRM environments, which is further resolved across intermediate component flexibilities in Fig.~\ref{NGP_diff_Kbr}. In the stiff BRP medium (red dashed line), the tracer exhibits a strict leptokurtic trend ($G(\tau) > 0$), despite the lower magnitude than in the flexible system, confirming that the branched architecture enforces persistent topological caging in the rigid environment. Conversely, in the stiff LRM medium (red solid line), well-separated linear and ring polymers provide open diffusive channels; consequently, the tracer dynamics are dominated by intermediate-time persistent motion, resulting in a predominantly platykurtic ($G(\tau) < 0$) profile.


\begin{figure*}
\includegraphics[width=14cm]{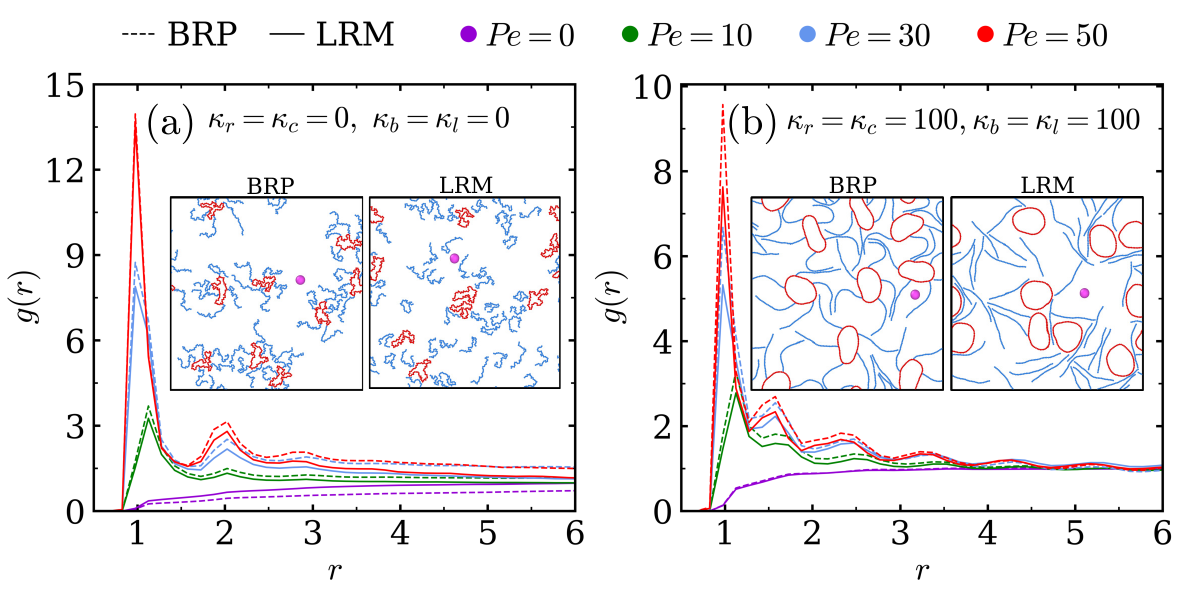}
\caption{Radial distribution function, $g(r)$, of polymer monomers around the tracer in the BRP (dashed lines) and LRM (solid lines) systems across varying propulsion strengths: $Pe = 0$ (violet), $10$ (green), $30$ (blue), and $50$ (red), for (a) fully flexible ($\kappa_c = \kappa_r = \kappa_b = \kappa_l = 0$) and (b) stiff ($\kappa_c = \kappa_r = \kappa_b = \kappa_l = 100$) polymer matrices. Insets display representative snapshots of the active tracer (magenta) and surrounding polymer conformations of the BRP and LRM systems.}
\label{RDF_Diff_Pe}   
\end{figure*}

\begin{figure}
    \includegraphics[width=8cm]{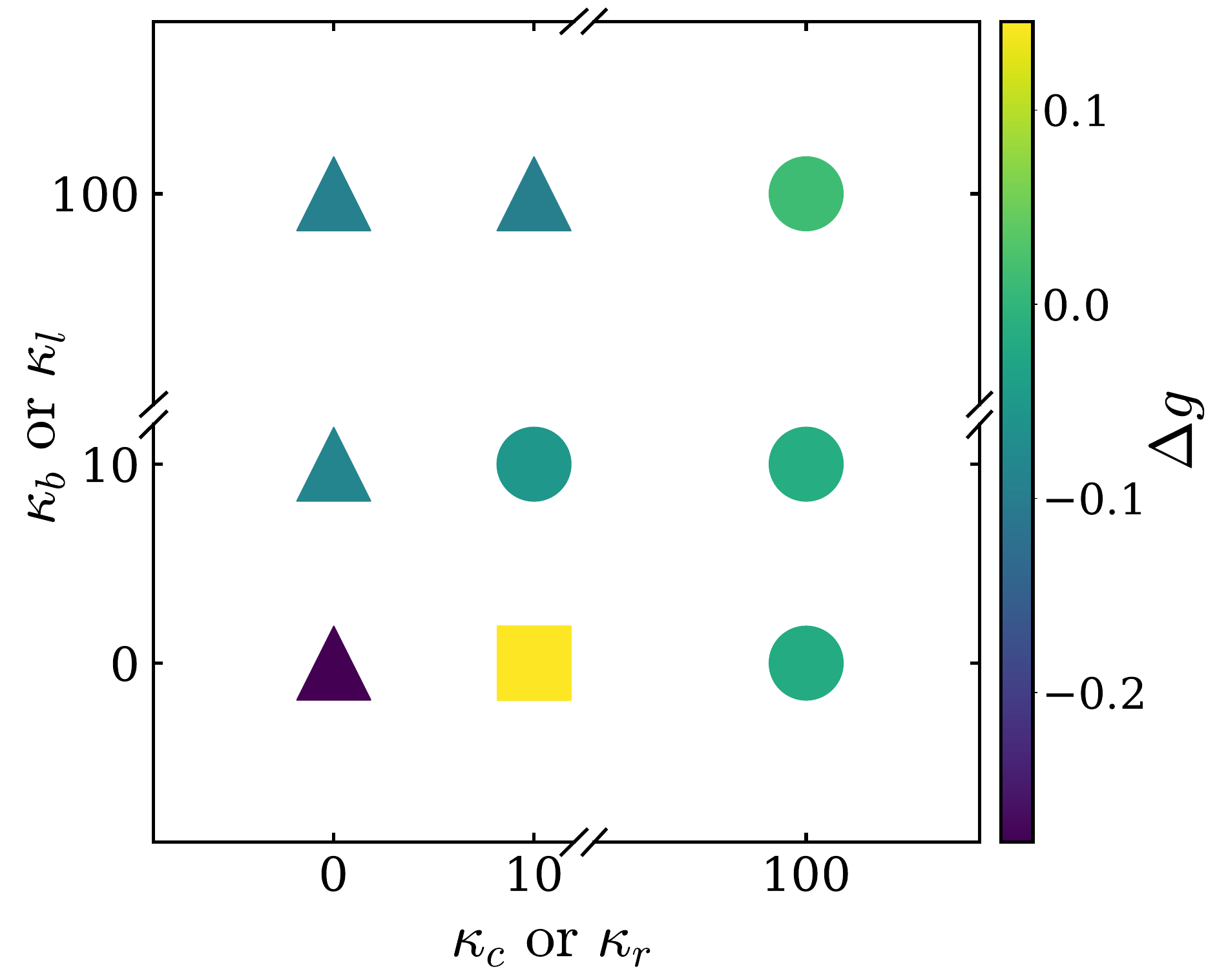}
  \caption{Phase diagram of the local monomer density contrast around a passive tracer ($Pe = 0$) across the component stiffness space $\{\kappa_b, \kappa_c\}$ (BRP) and $\{\kappa_l, \kappa_r\}$ (LRM). The color scale represents the radial distribution difference $\Delta g= g_{\text{BRP}}(r) - g_{\text{LRM}}(r)$ evaluated at $r = 3$, where triangles indicate $\Delta g < -0.1$, circles denote $-0.1 < \Delta g < 0.1$, and the square corresponds to $\Delta g > 0.1$.}
  \label{RDF_Contour}   
\end{figure}

\subsubsection{Radial Distribution Function, \texorpdfstring{$g(r)$}{g(r)}}

To quantify the spatial distribution of monomers around the tracer, we calculate the radial distribution function (RDF), $g(r)$ (Eq.~\ref{eq: RDF}). Fig.~\ref{RDF_Diff_Pe} shows $g(r)$ for the BRP and LRM systems across varying tracer propulsion strengths ($Pe$) in the flexible ($\kappa=0$) and stiff ($\kappa=100$) limits. The primary peak height increases with $Pe$ across the considered polymer flexibilities (Fig.~\ref{RDF_Diff_Pe}(a) \& (b)). Higher propulsion enables the active tracer to explore larger areas and interact with the polymers more frequently, resulting in a higher contact-peak intensity as it encounters more monomers. For active tracers ($Pe>0$), $g(r)$ consistently exhibits higher values in the BRP system compared to the LRM system across all flexibilities. The BRP matrix is a more structured, localized, and interconnected system. Due to this bond connectivity, a large number of monomers reside in the vicinity of the cyclic backbone. Consequently, active particles are trapped more frequently in the BRP environment, leading to a greater $g(r)$ amplitude (Fig.~\ref{RDF_Diff_Pe}, dashed lines). Notably, at the highest activity ($Pe=50$) in the flexible limit, the primary $g(r)$ amplitude becomes equal for both polymeric systems (Fig.~\ref{RDF_Diff_Pe}(a), red lines), a convergence that does not occur in the stiff limit (Fig.~\ref{RDF_Diff_Pe}(b), red lines). This higher peak in the stiff BRP matrix directly reflects the enhanced topological trapping of the tracer (Fig.~\ref{taMSD_Beta_diff_Pec}(c)).

However, for a purely passive particle ($Pe=0$) in a flexible system, the LRM exhibits a higher $g(r)$ value than the BRP system (Fig.~\ref{RDF_Diff_Pe}(a), violet lines). In the LRM, the passive tracer encounters more uniformly dispersed monomers; in contrast, branch connectivity localizes each BRP into a compact cluster, lowering the average local monomer probability around a thermal tracer (Fig.~\ref{RDF_Diff_Pe}(a), inset).

Polymer stiffness also dictates the local monomer distribution by altering the overall shape and size of the polymeric units (Fig.~\ref{Rg_ASP_n5_l30}). Peak magnitudes of $g(r)$ across all activities are higher in systems with $\kappa=0$ than in those with $\kappa=100$. Flexible polymers adopt a compact, coiled structure, allowing a larger number of monomers to lie within the local neighborhood of the tracer at any given time. Stiff polymers, on the contrary, exhibit an extended conformation (Fig.~\ref{RDF_Diff_Pe}(b), inset). This geometric shift is verified by the mean squared radius of gyration ($\langle R_{g}^2 \rangle$) and the mean asphericity ($\langle A \rangle$), both of which increase monotonically with stiffness $\kappa$ (Fig.~\ref{Rg_ASP_n5_l30}(a,  b)). These structural differences strongly affect the spatial distribution of the matrix. In the fully flexible limit, monomers in the LRM system are more spread out than in the BRP system (Fig.~\ref{RDF_Diff_Pe}(a), inset). In contrast, stiff polymers ($\kappa=100$) are more evenly distributed in both topologies (Fig.~\ref{RDF_Diff_Pe}(b), inset), leading to nearly identical $g(r)$ profiles over larger distances. Finally, the stiff matrix shifts the secondary peak to a much shorter distance ($r \sim 1.5$) compared to the flexible matrix ($r \sim 2$).

In particular, for a passive particle ($Pe = 0$), we observe a non-monotonic crossover in the relative magnitude of $g(r)$ between the BRP and LRM systems across different polymer stiffness conditions (Fig.~\ref{RDF_Diff_Pe_All_K},violet lines). To quantify this local monomer distribution, we evaluate the difference of $g(r)$ at $r=3$, i.e., $\Delta g$ = $g_{BRP}(r)- g_{LRM}(r)$, as shown in Fig. \ref{RDF_Contour}, where negative, $\Delta g < -0.1$ (triangles), positive $\Delta g > 0.1$ (square), and nearly equal $|\Delta g| < 0.1$, (circles) regimes delineate how backbone and branch rigidities jointly govern local monomer crowding around the tracer.


\begin{figure*}
\centering
\includegraphics[width=16cm]{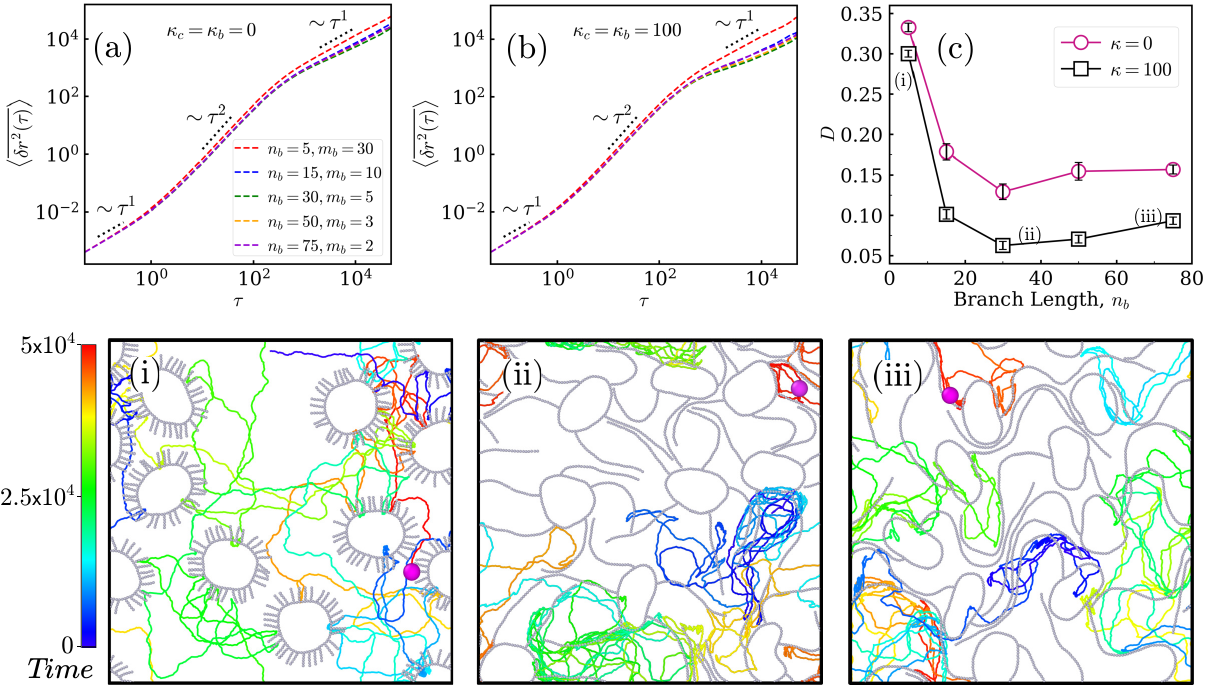}
\caption{Ensemble-averaged time-averaged mean squared displacement (etMSD) of a highly active tracer ($Pe = 50$) in the BRP system across varying combinations of branch length ($n_b$) and branch number ($m_b$) at a fixed packing fraction, for (a) fully flexible ($\kappa_c = \kappa_b = 0$) and (b) stiff ($\kappa_c = \kappa_b = 100$) polymers. (c) Corresponding long-time diffusion constant, $D$, as a function of $n_b$, exhibiting a non-monotonic dependence in both flexibility limits. The bottom panels display representative conformations (gray) of stiff BRPs ($\kappa = 100$) at the three branching geometries marked in (c): (i) $n_b = 5\bb{,}~m_b = 30$, (ii) $n_b = 30\bb{,}~m_b = 5$, and (iii) $n_b = 75\bb{,}~m_b = 2$, overlaid with a single tracer trajectory (magenta) color-coded by time from $0$ (blue) to $5 \times 10^4$ (red). Note that the tracer is enlarged for better visualization; the calculations are done considering the same diameter as the monomers.}
\label{tamsd_diff_num_len}
\end{figure*}


\subsection{Effect of Branch Number and Lengths: Fixed Packing Fraction}

At a constant packing fraction, the total number of monomers $N_b=n_c+\tilde{n}_b$ in each BRP remains fixed, where $\tilde{n}_b=m_b \times n_b$ denotes the total number of branch monomers, $m_b$ is the number of branches per BRP, and $n_b$ is the branch length. Thus, for a fixed backbone length $n_c$, increasing $m_b$ requires a corresponding decrease in $n_b$ to maintain a constant $\tilde{n}_b$. While in the preceding sections $n_b = 30$ and $m_b = 5$, here we investigate the effect of varying $n_b$ and $m_b$ on tracer dynamics under a constant packing fraction to determine whether an optimal branching geometry exists for restricting tracer motion. We compute the etMSD of a highly active tracer ($Pe = 50$) across various BRP architectures under fully flexible ($\kappa=0$) and stiff ($\kappa=100$) limits (Fig.~\ref{tamsd_diff_num_len}). As shown in Fig.~\ref{tamsd_diff_num_len}(a), in the flexible system, the long-time etMSD initially decreases with increasing $n_b$, reaches a minimum at $n_b=30$ ($m_b=5$, green line), and subsequently increases for longer branches. This non-monotonic behavior persists in the stiff system (Fig.~\ref{tamsd_diff_num_len}(b)). Notably, for the shortest branch length ($n_b=5, m_b=30$, red line), the tracer exhibits normal long-time diffusion across both flexibilities, whereas it undergoes long-time anomalous subdiffusion ($\alpha<1$) for all other branch combinations in the stiff architecture (Fig.~\ref{tamsd_diff_num_len}(b)).

To further quantify this non-monotonic behaviour of diffusivity, we evaluate the long-time diffusion constant, $D$, as a function of branch length $n_b$ (Fig.~\ref{tamsd_diff_num_len} (c)). Across all branch lengths, $D$ is lower for $\kappa=100$ (black squares) than for $\kappa=0$ (pink circles), as the stiff polymer matrix restricts the tracer more severely. In both flexibility limits, $D$ varies non-monotonically with $n_b$, exhibiting three distinct structural regimes marked as (i), (ii), and (iii) in Fig.~\ref{tamsd_diff_num_len}(c). The bottom panels of Fig.~\ref{tamsd_diff_num_len} display the corresponding polymer conformations alongside a representative trajectory of the tracer (magenta), color-coded by its time evolution. For numerous short branches ((i) $n_b=5, m_b=30$), the branches remain tightly localized around the cyclic backbones, leaving wide interstitial channels for the tracer to diffuse freely. Conversely, for a few long branches ((iii) $n_b=75, m_b=2$), the sparse branching leaves large open gaps between chains. At the intermediate geometry ((ii) $n_b=30, m_b=5$), the branches span the inter-ring space effectively to form interconnected cages, strongly trapping the active particle. Hence, a branched ring polymer topology with \textit{five branches}, each comprising \textit{thirty monomers}, provides the optimal combination to constrain the equal-sized active Brownian particle.

\section{Summary and Conclusion}
\label{sec-conclusion}

In this work, we have investigated the transport of passive and active tracers in a branched ring polymer (BRP) environment, where linear branches are grafted onto a cyclic backbone. To isolate the effect of this intramolecular architecture from simple steric crowding, we benchmarked the dynamics against a reference mixture of unlinked linear and ring polymers (LRM), kept at the same packing fraction, branch length, and bending stiffness. We systematically varied the tracer activity from the passive ($Pe = 0$) to the highly active ($Pe = 50$) limit and the component rigidity from fully flexible ($\kappa = 0$) to stiff ($\kappa = 100$).

The passive tracer serves as a useful baseline. Its transport is essentially identical in flexible and stiff polymeric environments across both BRP and LRM matrices, and its long-time diffusion constant is largely insensitive to $\kappa$ (Fig.~\ref{D_Pe}). Driven solely by thermal fluctuations, without any intrinsic activity, the passive tracer only samples its immediate local surroundings and does not explore the extended spatial length scales required to probe the global macromolecular connectivity. The macromolecular architecture therefore stays hidden until the tracer is driven.

In contrast, active tracers clearly distinguish between the two topological environments. When both the branches and backbones are flexible, the difference between the two systems is relatively weak: the tracer follows the canonical sequence of active transport, normal diffusion at short times, superdiffusion at intermediate times, and a return to normal diffusion at long times. However, the two systems diverge strikingly when the polymer components are stiff. In the stiff LRM, the tracer returns to normal diffusion at long times, whereas in the stiff BRP matrix it undergoes pronounced long-time subdiffusion, with $\alpha(\tau)<1$ (Fig.~\ref{taMSD_Beta_diff_Pec}). Notably, this long-time subdiffusion intensifies with increasing activity.  A possible explanation of this intriguing transport is that more persistent trajectories of active tracers drive the tracer deeper inside the topological pockets formed by the rigid branches, from which escape becomes increasingly difficult.

The corresponding long-time diffusion constants further quantify this topological contrast (Fig.~\ref{D_Pe}). Across the entire range of activity and stiffness considered, the diffusion constant remains systematically lower in the BRP matrix than in the LRM. Moreover, the dependence on chain stiffness is qualitatively opposite in the two architectures: increasing $\kappa$ enhances diffusivity in the unlinked LRM, whereas in the BRP system, increasing branch stiffness produces a systematic reduction. This striking contrast demonstrates that active tracer transport is governed not simply by the overall stiffness or monomer density of the medium, but by the spatial organization and intramolecular connectivity of the polymeric matrix.

This topological caging in BRP system induces pronounced dynamical heterogeneity and transient weak ergodicity breaking. The ensemble-averaged and time-averaged MSDs exhibit a clear discrepancy at intermediate lag times (Fig.~\ref{All_MSD}), accompanied by a systematically higher ergodicity-breaking parameter in the BRP system than in the LRM baseline (Fig.~\ref{EB_Parameter}), reflecting stronger trajectory-to-trajectory fluctuations due to variable residence times inside the branched cages. This microscopic heterogeneity is further corroborated by the van Hove correlation function and the non-Gaussian parameter (Fig.~\ref{van_Hove} and \ref{NGP_diff_Pec}). In the stiff limit at high activity ($Pe = 50$), the long-time displacement distribution in the unlinked LRM remains nearly Gaussian with platykurtic intermediate-time dynamics ($G(\tau) < 0$), whereas the BRP matrix exhibits a sharply peaked, non-Gaussian distribution and a strictly positive, leptokurtic $G(\tau) > 0$. Together, these statistical signatures confirm that active transport in the BRP matrix is governed by the coexistence of persistent self-propelled motion and intermittent topological trapping.

We further quantified spatiotemporal tracer relaxation dynamics using the self-intermediate scattering function, $F_s(q,\tau)$. At high activity, the tracer exhibits a systematically slower structural relaxation (longer $\tau_s$) in the BRP system than in the LRM at small wave vectors ($q = 1$, Fig.~\ref{ISF_Beta_tauS}). Interestingly, this topological difference diminishes at larger wave vectors, with the relaxation curves for the two systems nearly overlapping at $q=10$ (Fig.~\ref{ISF_Kwv_1_10}). This length-scale wave-vector dependence confirms that topological trapping operates predominantly at large macroscopic length scales, whereas short-range dynamics is governed by local monomer collisions and remains largely insensitive to polymer connectivity. In both environments, active tracer relaxation follows a compressed exponential decay ($\beta > 1$). At high activity, the stretching exponent $\beta$ plateaus at a lower value in the BRP matrix than in the LRM, reflecting the frequent interruption of ballistic displacements by the branched cages. It shows a deviation from that in LRM, suggesting that the distinction between the two environments lies in the underlying relaxation mechanism.

Finally, this study unravels that branched ring polymers possess an optimal branching architecture for efficiently confining the active tracer. Since the total number of monomers per molecule is fixed, the number of branches and their length cannot be chosen independently, and increasing the branch number ($m_b$) necessarily reduces the branch length ($n_b$). Varying these parameters simultaneously at constant packing fraction reveals that tracer diffusivity depends non-monotonically on branch length\bb{:} the diffusion constant first decreases, reaches a minimum at $n_b=30$ and $m_b=5$, then subsequently increases (Fig.~\ref{tamsd_diff_num_len}). This non-monotonicity arises from the competing spatial distributions in the two limiting regimes. Numerous short branches ($n_b = 5, m_b = 30$) remain tightly localized around the cyclic backbone and fail to span the interstitial space between polymers, whereas a few long branches ($n_b = 75, m_b = 2$) leave large open voids through which the tracer can readily escape. Only the intermediate branching architecture ($n_b = 30, m_b = 5$) bridges the inter-ring space effectively to form an interconnected mesh, maximizing topological trapping and anomalous subdiffusion.

Overall, our results establish that monomer packing fraction alone is insufficient to characterize the effect of macromolecular crowding. Two systems at an identical packing fraction can exhibit qualitatively different transport regimes, and in a stiff branched-ring matrix, self-propulsion counterintuitively hinders long-time transport by driving the tracer into topological traps. While the present two-dimensional model captures the essential interplay between intramolecular connectivity, chain stiffness, and active propulsion, extending this framework to three-dimensional networks where ring threading and concatenation can occur, as well as incorporating hydrodynamic interactions, higher packing fractions, and direct pore-size distributions will provide further quantitative insights into active transport in complex biological and synthetic soft matter.

\section{Acknowledgements}
 We would like to thank MATRICS, ANRF (Project No. MTR/2022/000833) for financial support. The authors acknowledge the computational resources  PARAM Rudra and PARAM Shakti Cluster from the National Supercomputing Mission (NSM), Government of India.

\section{Acronyms}
\label{sec-acronyms}
\begin{itemize}
    \item BRP: Branch Ring Polymer 
    \item LRM: Linear Ring Mixture
    \item BP: Branched Polymer in BRP system
    \item CB: Cyclic Backbone in BRP system
    \item LP: Linear Polymer in LRM system
    \item RP: Ring Polymer in LRM system
    \item $\kappa_{b}$: Branch Polymer Stiffness
    \item $\kappa_{c}$: Cyclic Backbone Polymer Stiffness
    \item $\kappa_{l}$: Linear Polymer Stiffness
    \item $\kappa_{r}$: Ring Polymer Stiffness
    \item $\kappa$: when $\kappa_{b}=\kappa_{c}$ or $\kappa_{l}=\kappa_{r}$ in a specific system
    \item eMSD: Ensemble-averaged MSD
    \item tMSD: Time-averaged MSD
    \item etMSD: Ensemble and time-averaged MSD
    \item WEB: Weak Ergodicity Breaking
    \item VHCF: van-Hove Correlation Function
    \item ISF: Self-intermediate Scattering Function
    \item NGP: Non-Gaussian Parameter
    \item RDF: Radial Distribution Function \\
\end{itemize}

\bibliographystyle{unsrt}
\section*{References}
\bibliography{BranchedRingPolymer.bib}

\clearpage

\onecolumngrid

\appendix

\counterwithin{figure}{section}

\centerline{\fontsize{14}{18}\selectfont\bfseries SUPPLEMENTARY INFORMATION}

\section{Trajectory Snapshots ($Pe=50$)} 
\subsection{Branched Ring Polymer (BRP)}
\begin{figure}[htp]
\includegraphics[width=16cm]{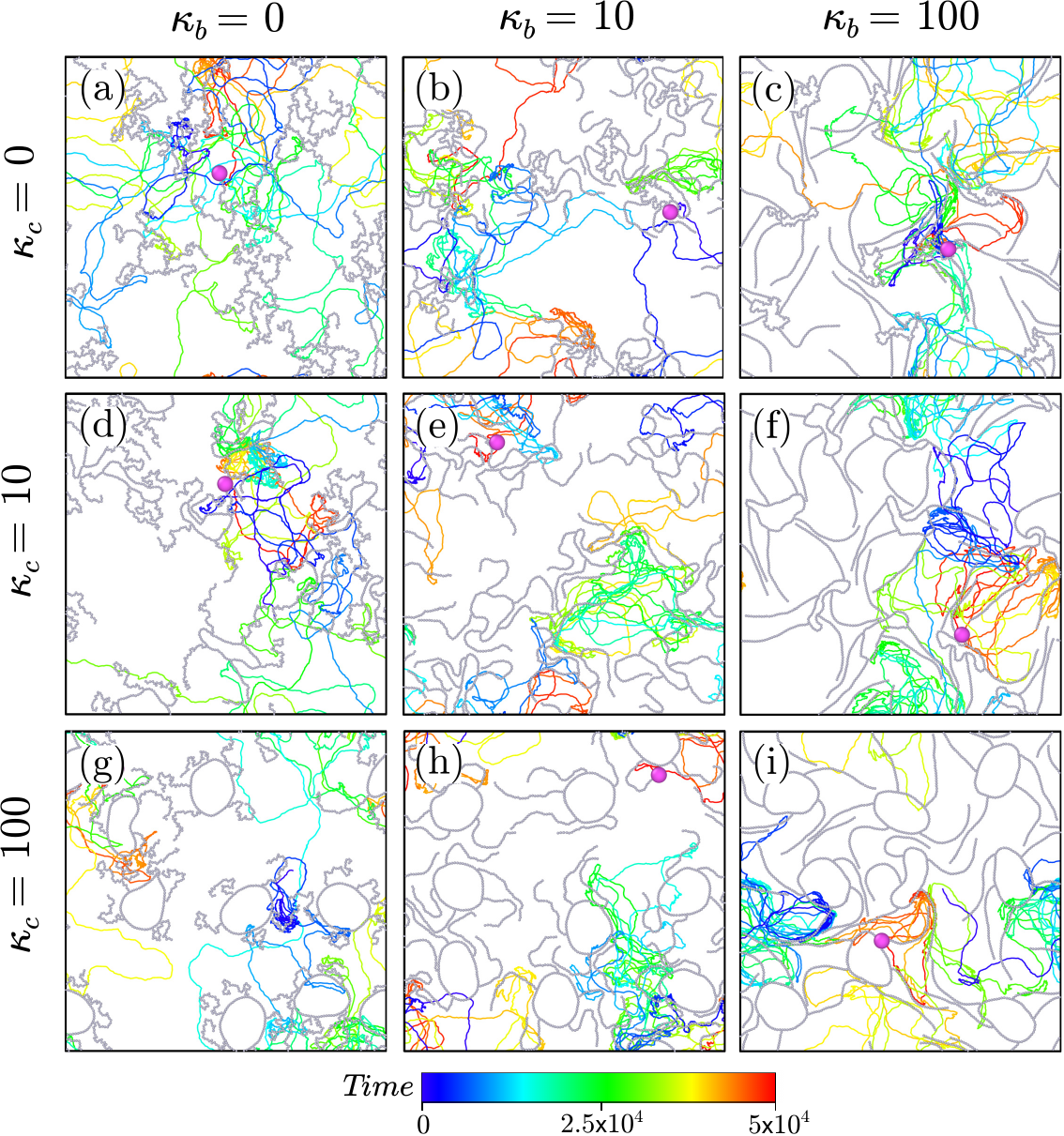}
\caption{Snapshots of polymer conformations (gray) and a representative active tracer ($Pe = 50$, magenta) trajectory across component stiffnesses in the Branched Ring Polymer (BRP) system. The cyclic backbones (CB) are fully flexible ($\kappa_c = 0$, top row: (a)-(c)), semiflexible ($\kappa_c = 10$, middle row: (d)-(f)), and stiff ($\kappa_c = 100$, bottom row: (g)-(i)), while the grafted branches vary from flexible ($\kappa_b = 0$, left column) to semiflexible ($\kappa_b = 10$, middle column) and stiff ($\kappa_b = 100$, right column). The trajectory color bar denotes time evolution in a single realization from $0$ (blue) to $5 \times 10^4$ (red). For visual clarity, the tracer is depicted larger than its actual monomer size. The simulation videos related to $\kappa_r=\kappa_l=0$ and $\kappa_r=\kappa_l=100$ are shown in supplementary movies (Movie S3 \& Movie S7).}
\label{BRP_Conf_Trj}
\end{figure}
\clearpage
\subsection{Linear Ring Mixture (LRM)}
\begin{figure}[htp]
\includegraphics[width=16cm]{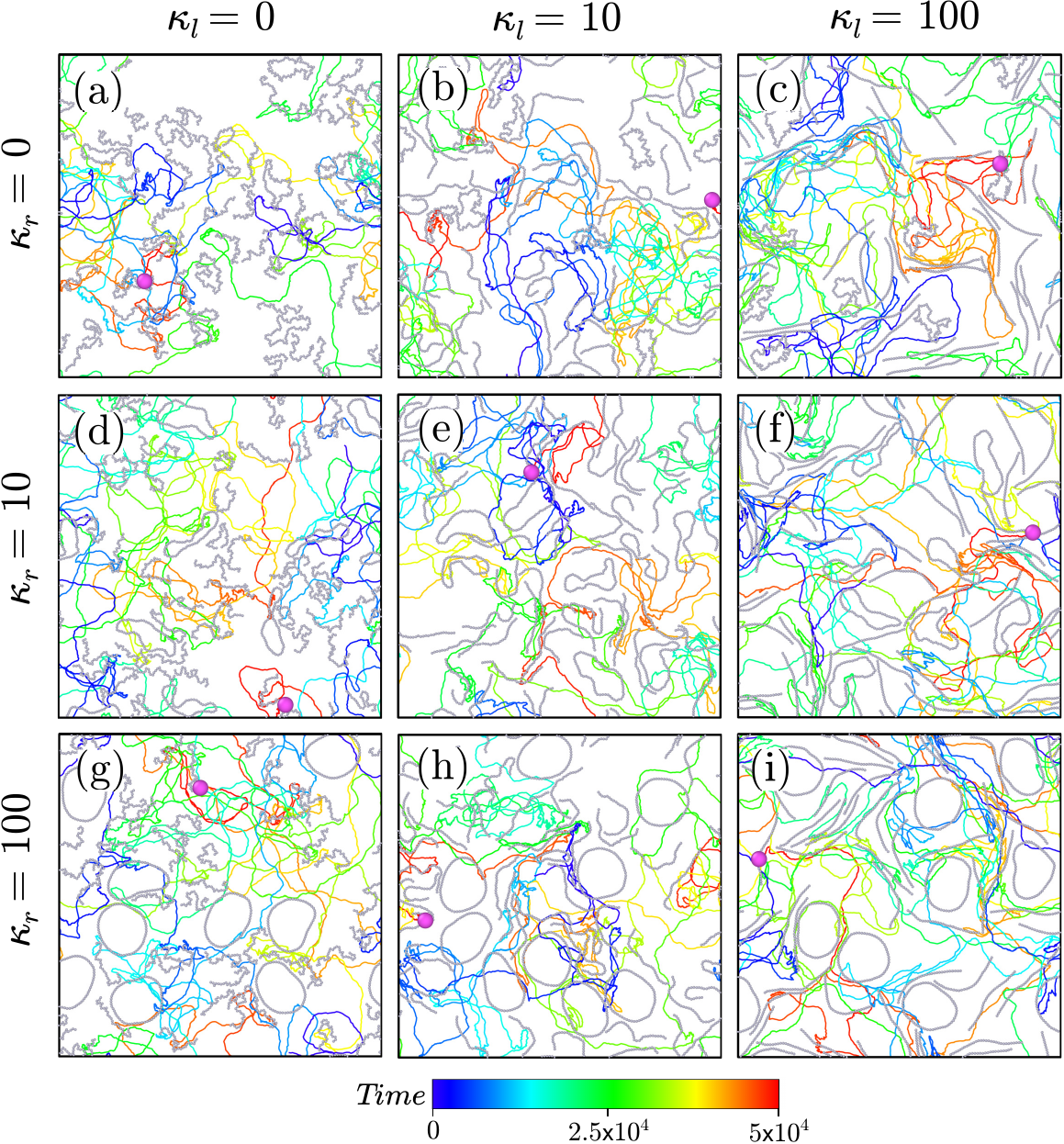}
\caption{Snapshots of polymer conformations (gray) and a representative active tracer ($Pe = 50$, magenta) trajectory across component stiffnesses in the unlinked Linear-Ring Mixture (LRM) system. From the top row to the bottom row, the ring polymer (RP) stiffness increases ($\kappa_r = 0, 10, 100$), while the linear polymer (LP) stiffness increases from left to right ($\kappa_l = 0, 10, 100$). The color-coded curve traces the path traversed by the tracer from $t = 0$ (blue) to $5 \times 10^4$ (red). The tracer is shown at a larger scale for better visibility, although its actual size is identical to that of the monomers. The simulation videos correspending to $\kappa_r=\kappa_l=0$ and $\kappa_r=\kappa_l=100$ are presented in supplementary movies(Movie S4 \& Movie S8).}
\label{LRM_Conf_Trj}
\end{figure}
\clearpage
\section{Mean Squared Displacement}
\subsection{Time Averaged Mean Squared Displacement and its Logarithmic Exponent}
\begin{figure}[htp]
\includegraphics[width=\textwidth]{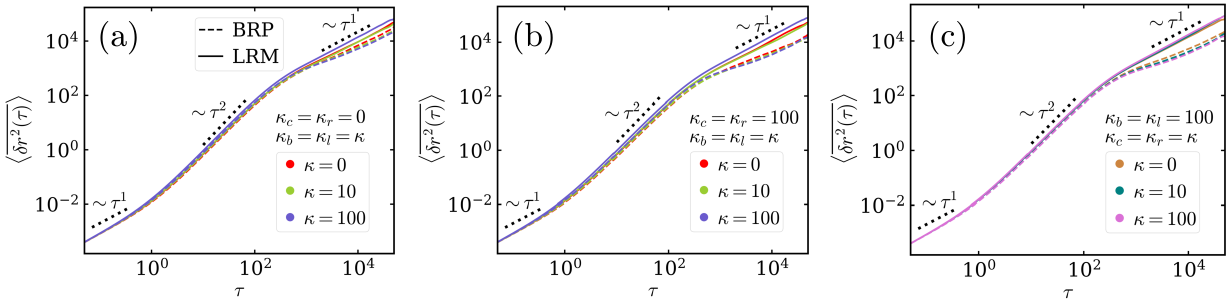}
\caption{Log-log plot of etMSD, $\langle\overline{\delta r^2(\tau)}\rangle$, versus lag time ($\tau$) for a highly active tracer ($Pe = 50$) in the BRP (dashed lines) and reference LRM (solid lines) systems. The cyclic backbones and rings are fixed at (a) flexible ($\kappa_c = \kappa_r = 0$) and (b) stiff ($\kappa_c = \kappa_r = 100$) limits, while the branch and linear chain stiffnesses ($\kappa_b = \kappa_l = \kappa$) are varied across flexible ($\kappa = 0$), semiflexible ($\kappa = 10$), and stiff ($\kappa = 100$) regimes. In (c), the branches and linear chains are kept stiff ($\kappa_b = \kappa_l = 100$) while the backbone and ring stiffnesses are varied uniformly ($\kappa_c = \kappa_r = \kappa = 0, 10, 100$).}
\label{tamsd_diff_Kbr_Kr}
\end{figure}

\begin{figure}[htp]
\includegraphics[width=\textwidth]{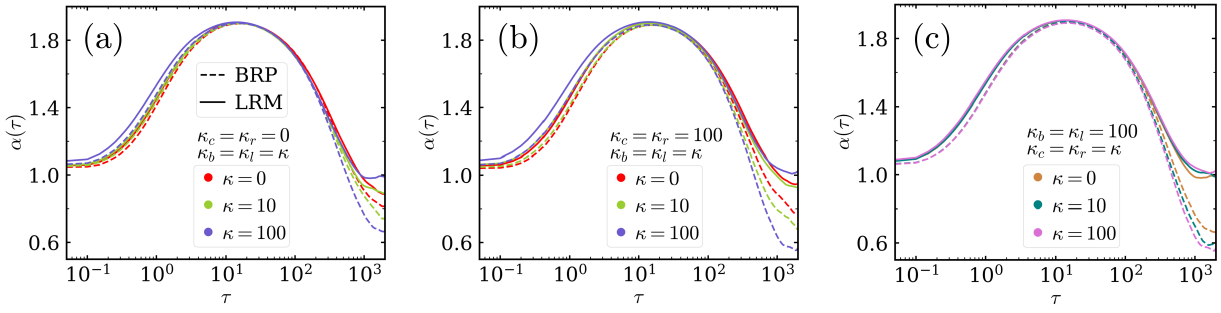}
\caption{Temporal variation of the local exponent, $\alpha(\tau)$, corresponding to the etMSD curves in Fig.~\ref{tamsd_diff_Kbr_Kr} for a highly active ABP ($Pe = 50$) in the BRP (dashed lines) and LRM (solid lines) systems. The cyclic backbones and rings are held at (a) flexible ($\kappa_c = \kappa_r = 0$) and (b) stiff ($\kappa_c = \kappa_r = 100$) limits while varying the branch and linear chain stiffnesses ($\kappa_b = \kappa_l = \kappa = 0, 10, 100$). In (c), $\kappa_b~\&~\kappa_l$ are fixed in the stiff limit ($\kappa_b = \kappa_l = 100$), while $\kappa_c$ and $\kappa_r$ are varied equally ($\kappa_c = \kappa_r = \kappa = 0, 10, 100$).}
\label{beta_diff_Kbr_Kr}
\end{figure}

Fig.~\ref{tamsd_diff_Kbr_Kr} shows the variation of the etMSD for a highly active tracer ($Pe=50$) across varying stiffnesses of the polymer components (cyclic backbones and branches in BRP; rings and linear chains in LRM). The stiffness of the cyclic backbone ($\kappa_c$) in BRP and that of the ring polymers ($\kappa_r$) in LRM play contrasting roles in the tracer dynamics. In the BRP matrix, long-time subdiffusive behavior is significantly stronger when the backbone is stiff ($\kappa_c=100$ ) compared to when it is flexible ($\kappa_c=0$ ), whereas the LRM system exhibits normal long-time diffusion irrespective of $\kappa_r$ (Fig.~\ref{tamsd_diff_Kbr_Kr}(a) \& (b) ). Stiff branches ($\kappa_b=100$) attached to a stiff backbone ($\kappa_c=100$) create a larger number of persistent local traps than those grafted onto a flexible backbone ($\kappa_c=0$). Hence, the tracer exhibits stronger subdiffusive behavior in the stiff BRP architecture (dashed blue line) , whereas increasing polymer stiffness in the unlinked LRM promotes tracer mobility, resulting in enhanced diffusion (solid blue line). Furthermore, when the branches in BRP and linear polymers in LRM are held stiff ($\kappa_b=\kappa_l=100$), tracer dynamics in the LRM remain largely unaffected by changes in $\kappa_r$, whereas in the BRP matrix, increasing $\kappa_c$ significantly suppresses long-time transport (Fig.~\ref{tamsd_diff_Kbr_Kr}(c)). The corresponding logarithmic exponent, $\alpha(\tau)$, provides further quantitative evidence of these distinct long-time diffusion regimes (Fig.~\ref{beta_diff_Kbr_Kr}). While $\alpha(\tau)$ remains close to unity in the LRM, reflecting normal diffusion, it falls well below unity ($\alpha<1$) in the BRP system and decreases further with increasing component stiffness, confirming the topologically driven subdiffusive behavior observed in Fig.~\ref{tamsd_diff_Kbr_Kr}.

\clearpage

\subsection{Mean Squared Displacement at $Pe=50$}
\begin{figure}[htp]
\includegraphics[width=\textwidth]{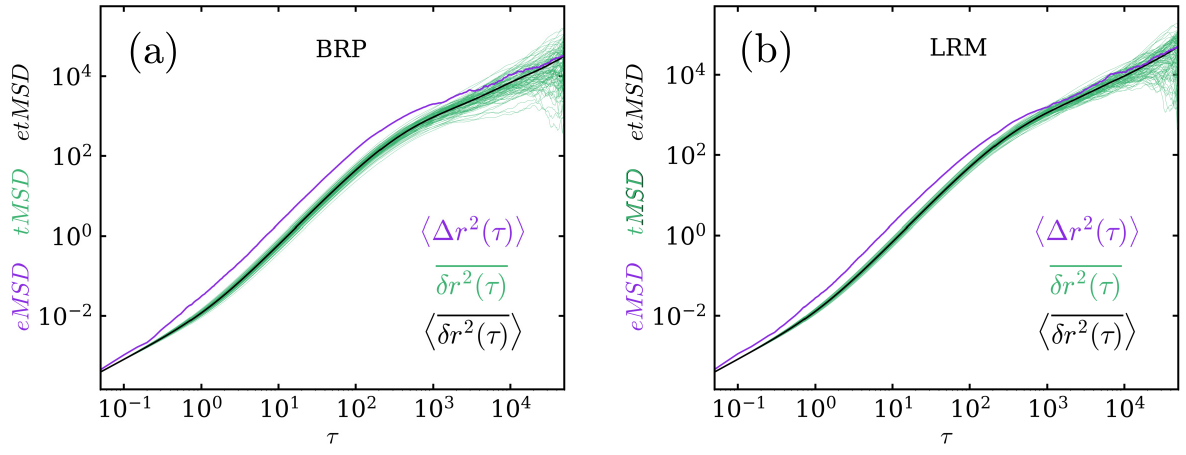}
\caption{Comparison of the eMSD $\langle\Delta r^2(\tau)\rangle$ (violet), individual tMSDs $\overline{\delta r^2(\tau)}$ (green), and etMSD $\langle\overline{\delta r^2(\tau)}\rangle$ (black) for a highly active tracer ($Pe = 50$) in fully flexible polymeric environments: (a) the BRP system ($\kappa_c = \kappa_b = 0$) and (b) the reference LRM system ($\kappa_r = \kappa_l = 0$).} 
\label{all_MSD_K_0}
\end{figure}
\clearpage
\section{Ergodicity Breaking Parameter}
\begin{figure}[htp]
\centering
\includegraphics[width=14cm]{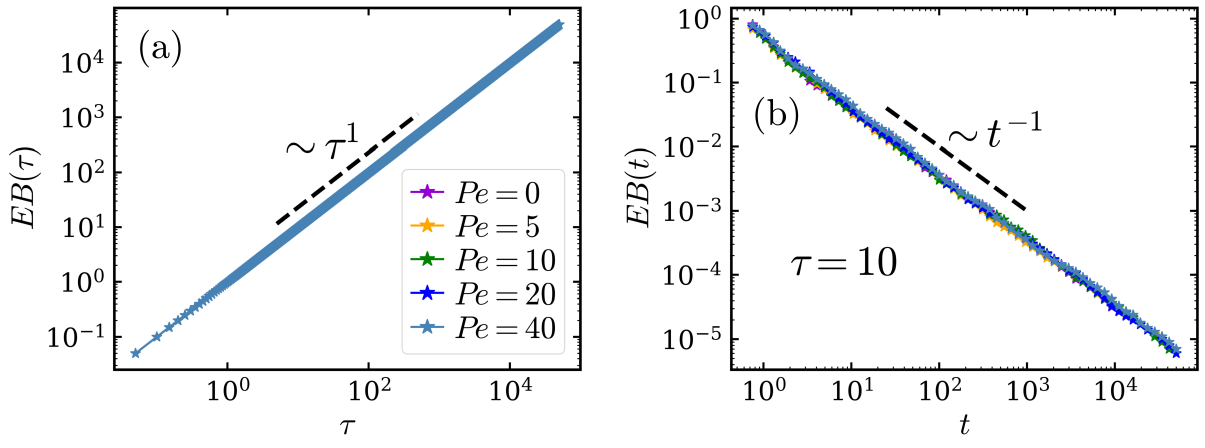}
\caption{Ergodicity breaking parameter, $EB$, of an active Brownian particle in unconfined (free) space as a function of (a) lag time ($\tau$) and (b) measurement time ($t$) at fixed lag time $\tau = 10$ for various propulsion strengths ($Pe$). The black dashed lines indicate the asymptotic power-law scalings with slopes (a) $EB \sim \tau^1$ and (b) $EB \sim t^{-1}$.}
\label{EBP_freespace}
\end{figure}

\begin{figure}[htp]
\centering
\includegraphics[width=16cm]{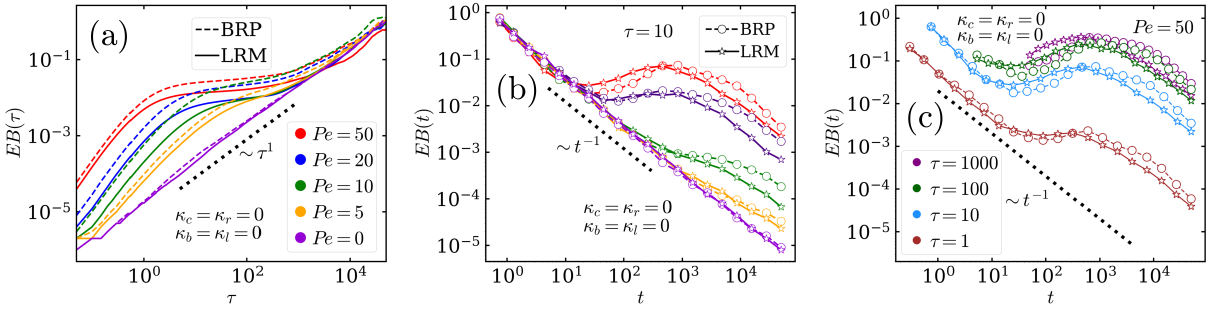}
\caption{Ergodicity breaking parameter, $EB$, in fully flexible polymer matrices ($\kappa_c = \kappa_r = \kappa_b = \kappa_l = 0$) for the BRP (dashed lines) and LRM (solid lines) systems, as a function of (a) lag time ($\tau$) and (b) measurement time ($t$) at a fixed lag time $\tau = 10$, for varying tracer propulsion strengths ($Pe = 0, 5, 10, 20, 50$). (c) $EB(t)$ versus measurement time across different lag times ($\tau = 1, 10, 100, 1000$) for a highly active tracer ($Pe = 50$). The thick black dotted line indicates the asymptotic scaling in (a) $EB \sim \tau^1$ and in (b, c) $EB \sim t^{-1}$.}
\label{EB_time_Diff_Pe}
\end{figure}

For an ABP in free space, $EB$ displays linear scaling with lag time, $EB(\tau) \sim \tau^1$ (Fig.~\ref{EBP_freespace}(a)), whereas it decreases inversely with measurement time, $EB(t) \sim t^{-1}$ (Fig.~\ref{EBP_freespace}(b)), across all activities ($Pe$). The same scaling pattern is observed in the presence of either polymer architecture for a passive tracer ($Pe=0$, violet lines in Fig.~\ref{EB_time_Diff_Pe}). Trajectory-to-trajectory fluctuations in the tMSD data increase with $\tau$ due to fewer independent averaging windows, leading to larger $EB(\tau)$ values. Conversely, over longer measurement times $t$, the time-averaged motion becomes less dependent on transient trapping history, producing less dispersion in the tMSD and a systematic decay in $EB(t)$. For active tracers ($Pe>0$), as shown in Fig.~\ref{all_MSD_K_0}, the intermediate-time mismatch between the etMSD and eMSD elevates $EB(\tau)$. This sustained deviation between the MSDs produces an intermediate-time plateau in Fig.~\ref{EB_time_Diff_Pe}(a) and a corresponding non-monotonic peak in $EB(t)$ at intermediate measurement times (Fig.~\ref{EB_time_Diff_Pe}(b, c)) for both systems with flexible polymers ($\kappa_c=\kappa_r=\kappa_b=\kappa_l=0$). Across all activities and lag times, ergodicity breaking remains systematically stronger in the BRP matrix than in the LRM system.

\clearpage
\section{Self-intermediate Scattering Function}
\begin{figure}[htp]
\centering
\includegraphics[width=10cm]{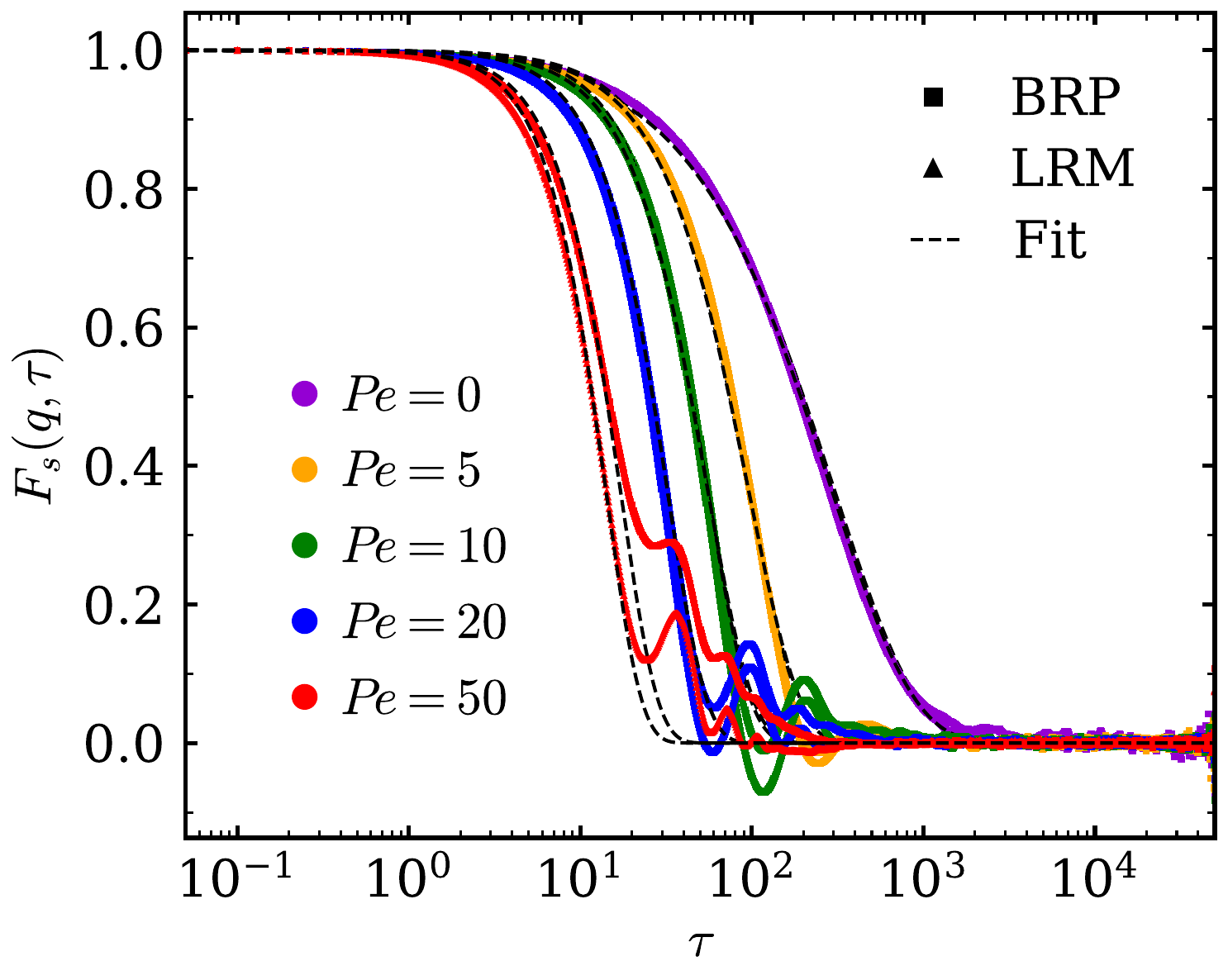}
\caption{Self-intermediate scattering function, $F_s(q,\tau)$, at wave vector $q = 1$ plotted against lag time $\tau$ for various propulsion strengths ($Pe$) of the tracer in the stiff BRP (squares) and LRM (triangles) systems ($\kappa_c = \kappa_b = \kappa_r = \kappa_l = 100$). The thin black dashed curves represent compressed-exponential fits according to Eq. $22$.}
\label{ISF_Fit}
\end{figure}
\begin{figure}[htp]
\centering
\includegraphics[width=16cm]{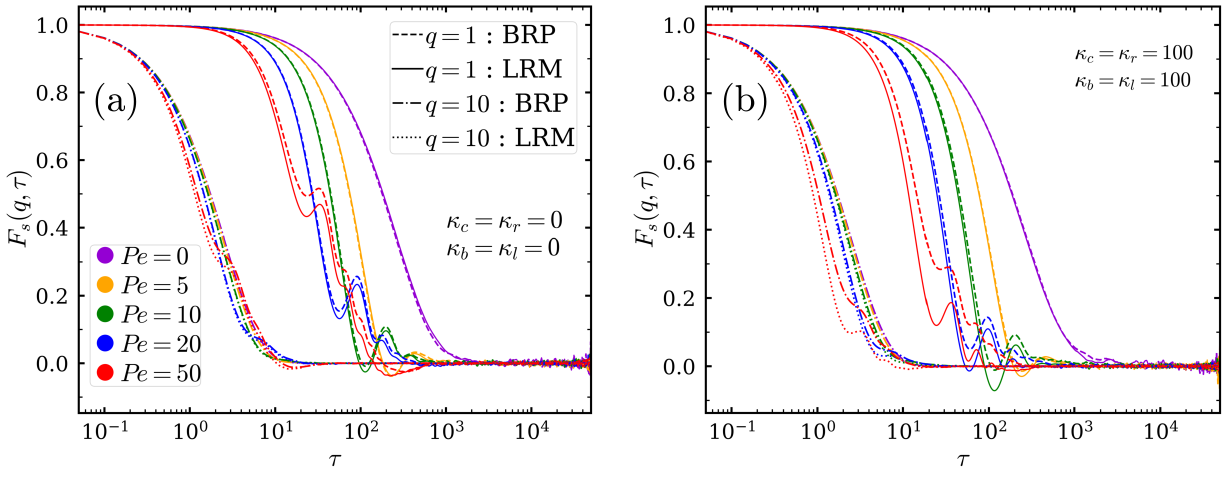}
\caption{Log-linear plot of the self-intermediate scattering function, $F_s(q,\tau)$, comparing large ($q = 1$) and short ($q = 10$) spatial length scales in (a) fully flexible ($\kappa_c = \kappa_b = \kappa_r = \kappa_l = 0$) and (b) stiff ($\kappa_c = \kappa_b = \kappa_r = \kappa_l = 100$) polymer matrices across varying tracer activities ($Pe$). Here dashed (BRP) and solid (LRM) lines represent $q=1$, whereas dotted dashed (BRP) and dotted (LRM) lines correspond to $q=10$.}
\label{ISF_Kwv_1_10}
\end{figure}
\clearpage
\section{Non-Gaussian Parameter}
\begin{figure}[htp]
\centering
\includegraphics[width=10cm]{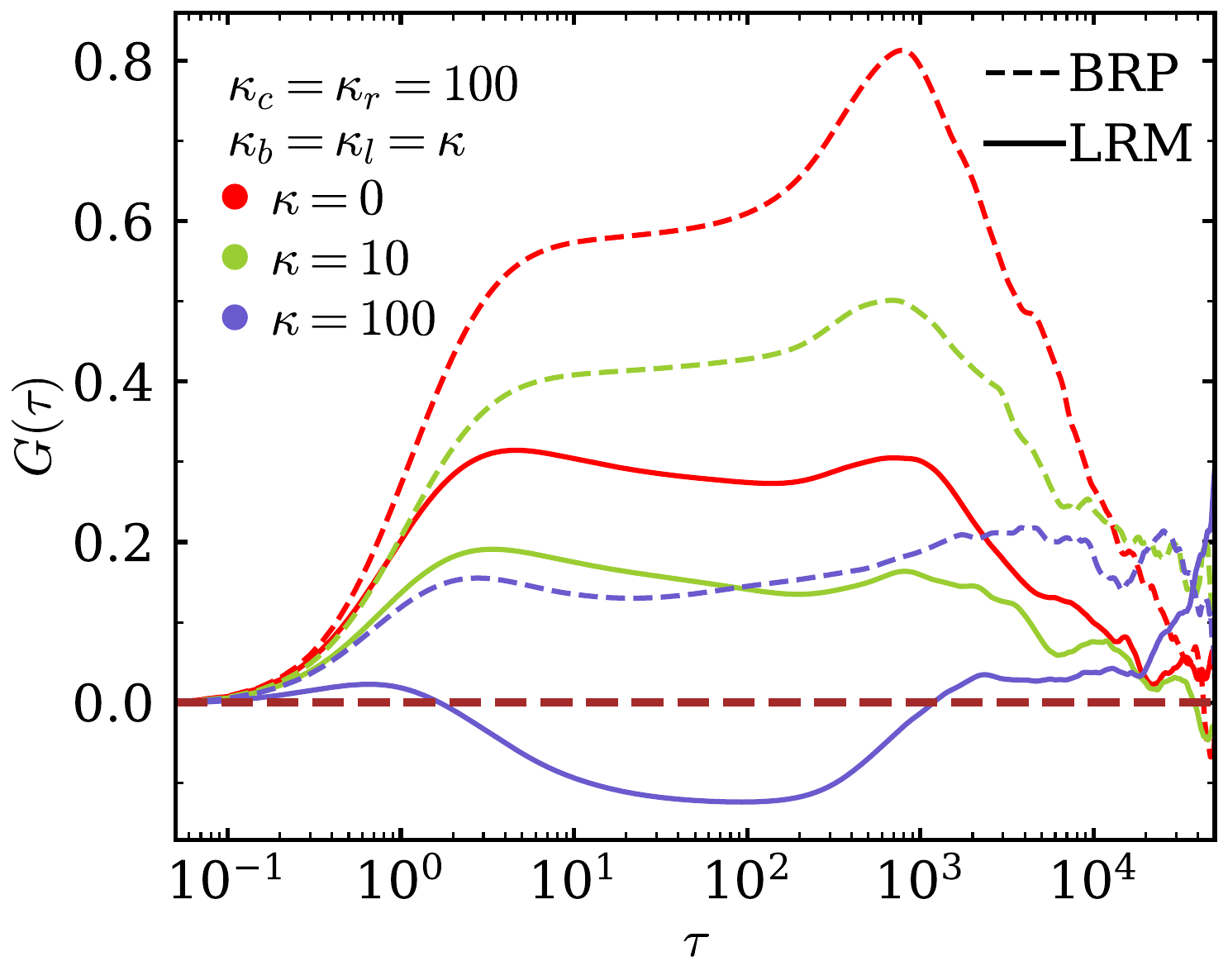}
\caption{Time evolution of the non-Gaussian parameter, $G(\tau)$, for a highly active ($Pe = 50$) tracer in the BRP (dashed lines) and LRM (solid lines) systems with stiff cyclic backbones and rings ( $\kappa_c = \kappa_r = 100$), while the stiffness of the branches and linear polymers ($\kappa_b = \kappa_l = \kappa$) is varied from flexible ($\kappa = 0$) to semiflexible ($\kappa = 10$) and stiff ($\kappa = 100$). The thick brown dashed horizontal line marks the baseline for Gaussian behavior, $G(\tau) = 0$.} 
\label{NGP_diff_Kbr}
\end{figure}

The time evolution of the non-Gaussian parameter, $G(\tau)$, for a highly active tracer ($Pe=50$) at fixed backbone and ring stiffness ($\kappa_c = \kappa_r = 100$) across various stiffnesses of the branch and linear polymers ($\kappa_b = \kappa_l = \kappa$) is shown in Fig.~\ref{NGP_diff_Kbr}. In both systems, $G(\tau)$ depends strongly on $\kappa$, especially in the intermediate time regime, with its positive magnitude decreasing monotonically with increasing $\kappa$. Across all values of $\kappa$, the leptokurtic deviation from Gaussianity ($G(\tau)=0$) is systematically greater in the BRP matrix compared to the LRM system. Increasing chain stiffness suppresses conformational fluctuations of the polymers and reduces local steric entanglement, thereby lowering $G(\tau)$. Notably, in the stiff limit ($\kappa = 100$), $G(\tau)$ remains strictly positive in the BRP matrix due to persistent topological caging, whereas it becomes negative (platykurtic) at intermediate times in the unlinked LRM as the tracer undergoes unhindered persistent motion.

\clearpage
\section{Radial Distribution Function}
\begin{figure}[htp]
\centering
\includegraphics[width=\textwidth]{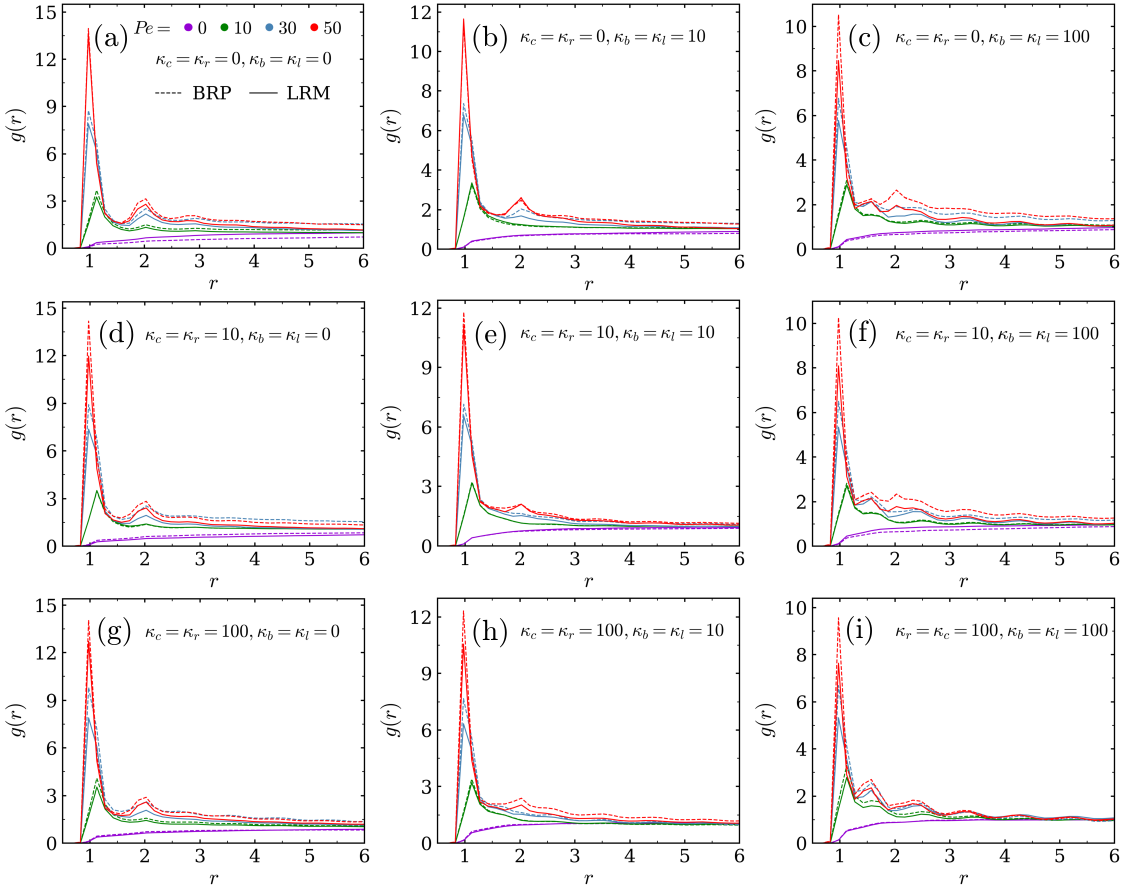}
\caption{Radial distribution function, $g(r)$, of monomers around the tracer at varying activities ($Pe = 0, 10, 30, 50$) in the BRP (dashed lines) and LRM (solid lines) systems across all combinations of component stiffness. The cyclic backbones (CB) in BRP and rings (RP) in LRM are fully flexible ($\kappa_c = \kappa_r = 0$, top row: (a)-(c)), semiflexible ($\kappa_c = \kappa_r = 10$, middle row: (d)-(f)), and stiff ($\kappa_c = \kappa_r = 100$, bottom row: (g)-(i)), while from left to right the branches (BP) and linear polymers (LP) vary from flexible ($\kappa_b = \kappa_l = 0$, left column) to semiflexible ($\kappa_b = \kappa_l = 10$, middle column) and stiff ($\kappa_b = \kappa_l = 100$, right column). The color scheme follows Fig. 9.}
\label{RDF_Diff_Pe_All_K}
\end{figure}
\clearpage
\section{Radius of Gyration and Asphericity}
\begin{figure}[htp]
\centering
\includegraphics[width=14cm]{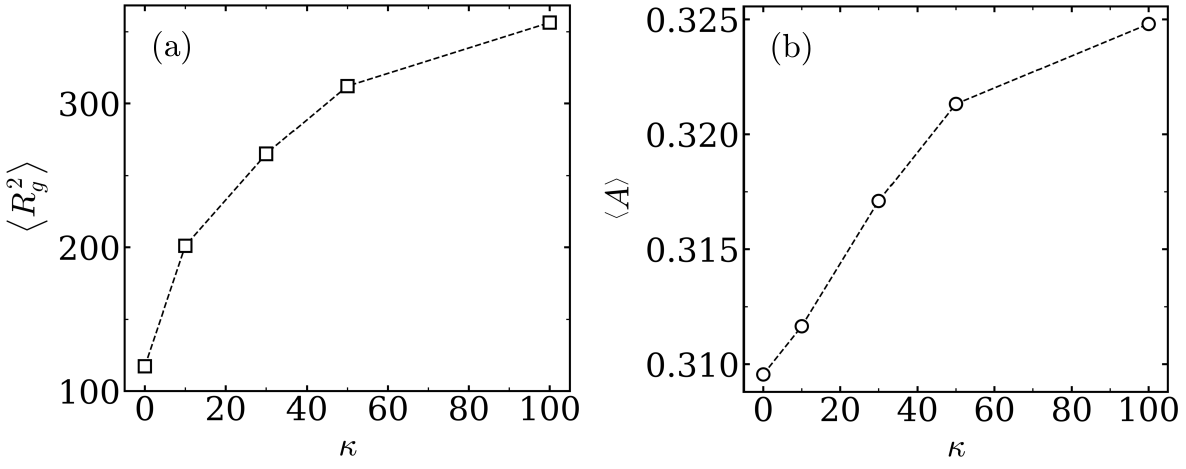}
\caption{Dependence of the overall size and shape of a branched ring polymer (BRP) on bending stiffness: (a) mean squared radius of gyration ($\langle R_g^2\rangle$) and (b) mean asphericity ($\langle A\rangle$) plotted as a function of the uniform component stiffness ($\kappa_c = \kappa_b = \kappa$). Both structural properties increase monotonically with $\kappa$, reflecting a transition from compact coils to highly extended and anisotropic conformations in the stiff limit. The dashed lines serve as guides to the eye.}
\label{Rg_ASP_n5_l30}
\end{figure}
\section{Movie Description}
\begin{itemize}
    \item {\textbf{Movie S1:} Dynamics of a passive tracer ($Pe=0$) in a fully flexible branched-ring polymer (BRP) matrix ($\kappa_c=\kappa_b=0$). The tracer undergoes conventional Brownian diffusion, sampling only its local surroundings within the flexible, topologically branched polymer environment.}
    \item {\textbf{Movie S2:} Brownian motion of a passive tracer ($Pe=0$) in the corresponding fully flexible linear-ring mixture (LRM) ($\kappa_r=\kappa_l=0$), providing a reference for thermal tracer transport in the unlinked polymeric environment.}
    \item {\textbf{Movie S3:} Dynamics of a highly active tracer ($Pe=50$) that undergoes persistent motion through the fully flexible BRP matrix ($\kappa_c=\kappa_b=0$), experiencing intermittent transient caging while} exploring the fluctuating branched polymer environment.
    \item {\textbf{Movie S4:} Active tracer ($Pe=50$) transport in the fully flexible LRM system ($\kappa_r=\kappa_l=0$), showing persistent exploration in the absence of interconnected topological constraints.}
    \item {\textbf{Movie S5:} Increasing the stiffness of the BRP constituents to ($\kappa_c=\kappa_b=100$ restricts the motion of a passive tracer ($Pe=0$), illustrating the local crowding effect imposed by the stiff polymer segments.}
    \item {\textbf{Movie S6:} In the stiff LRM environment ($\kappa_r=\kappa_l=100$), the passive tracer ($Pe=0$) experiences similarly restricted local transport due to the rigidity of the surrounding polymer matrix, providing a baseline comparison with the stiff BRP system.}
    \item {\textbf{Movie S7:} At high activity ($Pe=50$), the tracer moves persistently through the stiff BRP matrix ($\kappa_c=\kappa_b=100$) but undergoes prolonged localization within the interconnected pockets formed due to the branches, highlighting the emergence of severe topological caging.}
    \item {\textbf{Movie S8:} The corresponding highly active tracer ($Pe=50$) in the stiff LRM system ($\kappa_r=\kappa_l=100$) retains extended spatial exploration through open diffusive channels, in stark contrast to the pronounced localization observed in the stiff BRP matrix.}
\end{itemize}


\end{document}